%% file: main.tex
\documentclass[10pt, conference, compsocconf]{IEEEtran}

\usepackage{delarray}
\usepackage{boxedminipage}
\usepackage{multirow}
\usepackage{hhline}
\usepackage{ulem}
\usepackage{dcolumn}
\usepackage{amsmath}
\usepackage{caption, cite}
\usepackage{floatflt,epsfig,stfloats}
\usepackage{array}
\usepackage{graphicx,subfigure}
\usepackage{pifont}
\usepackage{setspace}
\usepackage{color}
\usepackage{fancyhdr}
\usepackage{xcolor}
\usepackage{soul}
\usepackage{hyperref}
\hypersetup{colorlinks, citecolor=Blue}
\soulregister\cite7
\soulregister\ref7
\soulregister\pageref7

\renewcommand{\theenumi}{\roman{enumi}}
\def\eg{{\em e.g.}, }
\def\ie{{\em i.e.}, }

\begin{document}

\title{R3-DLA (Reduce, Reuse, Recycle): A More Efficient Approach to Decoupled
Look-Ahead Architectures}

\author{\IEEEauthorblockN{Sushant Kondguli and Michael Huang} 
		\IEEEauthorblockN{Department of Electrical and Computer Engineering\\
						University of Rochester, Rochester, NY\\
						\{sushant.kondguli, michael.huang\}@rochester.edu
						}
}

\maketitle

\thispagestyle{plain}
\pagestyle{plain}

\input{abstract}


\input{intro}
\input{related}
\input{arch}
\input{eval}
\input{conclusions}

\section*{Acknowledgment}
This work is support in part by NSF under grants
1514433 and 1722847, and by a gift from Huawei. 

\scriptsize
\bibliographystyle{plain}
\bibliography{main}

\small
\appendices
\input{appendix}
\input{app}

\end{document}

%% file: abstract.tex
\begin{abstract}

%FIXME: Main bottlenecks in achieveing high single thread performance and 
% how they have become even more severe in new architectures.
% Techniques to achieve high single thread performance and issues with them.
% Very high level idea of proposed solution and why it makes sense!

Modern societies have developed insatiable demands for more computation
capabilities. Exploiting implicit parallelism to provide automatic performance
improvement remains a central goal in engineering future general-purpose
computing systems. One approach is to use a separate thread context to
perform continuous look-ahead to improve the data and instruction supply to
the main pipeline. Such a decoupled look-ahead (DLA) architecture can be
quite effective in accelerating a broad range of applications in a relatively
straightforward implementation. It also has broad design flexibility as the
look-ahead agent need not be concerned with correctness constraints. In
this paper, we explore a number of optimizations that make the look-ahead
agent more efficient and yet extract more utility from it. With these
optimizations, a DLA architecture can achieve an average speedup of 1.4 over
a state-of-the-art microarchitecture for a broad set of benchmark suites,
making it a powerful tool to enhance single-thread performance.

\end{abstract}

%% file: intro.tex
\section{Introduction}

Modern societies have developed insatiable demands for more computation
capabilities. While in certain segments, delivering higher performance
via intense human labor (manual parallelization and performance tuning)
is justifiable, in many other situations, such effort is not necessarily
effective nor is it particularly efficient when the extra resources (energy
and loss of productivity) are properly accounted for. Automatic performance
improvement remains a central goal in engineering future general-purpose
computing systems~\cite{hwu.micro06}. After all, the systems have always
been designed to increase automation and productivity and to free human from
drudgery (including debugging a parallel code).

\begin{figure}[htb]
    \centering
    \includegraphics[width=3in]{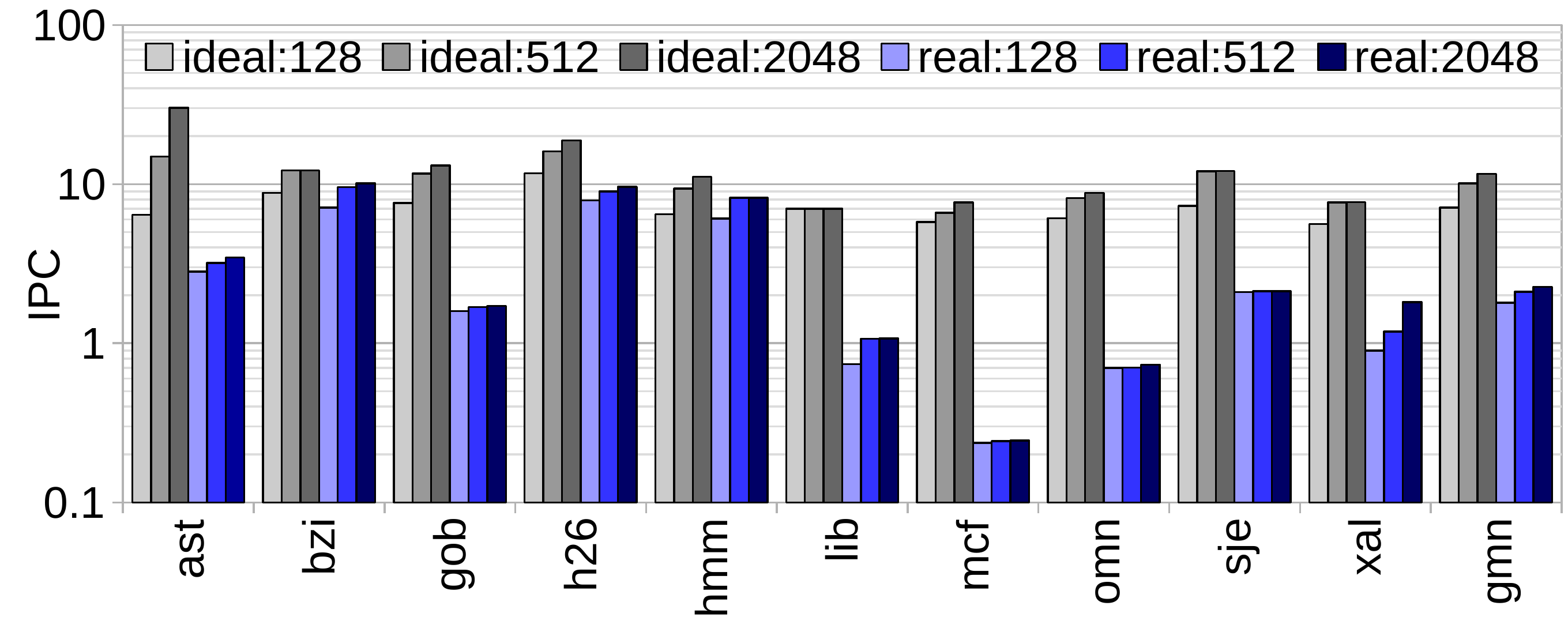}
	\caption{Implicit parallelism of integer applications. The three bars on the
	left indicate the amount of available parallelism (instructions per cycle)
	in the application when inspected with a moving window of 128, 512, or 2048
	instructions. The three bars on the right repeat the same experiments only
	this time under a realistic branch misprediction and cache miss situation
	that further constrains the available parallelism to be exploited. Note that
	the vertical axis is in log scale. \label{fig:ilp}} 
\end{figure}

The two traditional drivers for single-thread performance (faster cycles
and advancements in microarchitecture) have all but stopped in recent years
-- and for good reasons, since further gains from these approaches will
come at significant costs. However, the level of implicit parallelism is
quite high even in non-numerical codes. The real question is whether we can
realize the potential without undue costs. The typical monolithic out-of-order
microarchitecture appears to have significant challenges exploiting this level
of parallelism. In particular, the instruction and data supply subsystem
shoulders a significant responsibility. Fig.~\ref{fig:ilp} illustrates this
point. When data and instruction supply subsystem is idealized, the average
level of implicit parallelism is significantly higher (about 5x) than when it
is not, suggesting the subsystem as a target for improvements.

One possible way of strengthening this subsystem is to use a decoupled
look-ahead (DLA) architecture. In DLA, a self-sufficient thread guides the
look-ahead activities largely independent of the actual program execution.
Simply having a thread for continuous look-ahead is not enough. Many elements
have to function properly and efficiently to create a system that can offer
sustained, deep, and high-quality look-ahead that will provide significant
benefits at acceptable costs. In this paper, we discuss four optimizations
on top of a basic DLA architecture. Their effects are mainly to increase the
efficiency of the look-ahead thread and simultaneously extract more utility
out of it. The spirit of techniques are somewhat analogous to the ``reduce,
reuse, and recycle" mantra in waste reduction, hence the name R3-DLA.

The rest of the paper is organized as follows: Sec.~\ref{sec:related} discusses
variants of DLAs and the related concepts of helper threads;
Sec.~\ref{sec:design}
explains the design details of the proposed optimizations; Sec.~\ref{sec:eval}
performs experimental analysis on these optimizations; and finally
Sec.~\ref{sec:conclusions} summarizes the findings.

%% file: related.tex
\section{Background and Related Works}
\label{sec:related}

A basic on-demand caching system is only part of the solution to a
high-performance instruction and data supply subsystem. Anticipating
needed data and prefetching them is an indispensable component. 
Canonical prefetchers hard-wire expected access patterns and search for clues that
suggest the start of targeted access patterns. Stride prefetchers are
perhaps the first example that comes to mind. They target access streams
that have a constant stride and is comparatively easy to identify and
track~\cite{jouppi.isca90, palacharla.isca94, hur.micro06, ishii.jilp11,
pugsley.hpca14, michaud.hpca16, kondguli.iccd17, kondguli.isca18}. Other prefetchers target
memory access properties such as spatial locality~\cite{somogyi.isca06, chen.hpca04,
burger.iccd01, kumar.isca98, dimitrov.jilp11, cantin.asplos06, somogyi.isca09}
and temporal locality~\cite{joseph.isca97, 
jain.micro13, somogyi.isca09, chou.micro07} to prefetch
data. Correlating address streams~\cite{nesbit.hpca04, wenisch.hpca09,
nesbit.pact04} has been shown to help in prefetching some hard to prefetch
data, albeit with lower accuracy. Some prefetching techniques also target
pointer chasing accesses~\cite{roth.asplos98, roth.isca99, jain.micro13,
collins.micro02, cooksey.asplos10, wang.can03, ebrahimi.hpca09}.

Ultimately, not all accesses can be described by simple address patterns.
Obtaining addresses through partial execution of the program represents
a broad class of prefetching approaches. On one extreme of the design
spectrum, many short threads are launched as helpers to precompute
information for data or instruction supply~\cite{ dubois.tr98, farcy.micro98,
roth.asplos98, roth.ics99, chappell.isca99, roth.hpca01, zilles.isca01,
annavaram.isca01, luk.isca01, collins.isca01, moshovos.ics01, collins.micro01,
wang.hpca02, kim.asplos02b, chappell.isca02, liao.pldi02, chappell.micro02,
chaudhry.ieeemicro05, chaudhry.isca09, collins.micro02, atta.micro15, 
kim.cgo04, kim.tocs04, wang.ieeemicro04, song.pact05, lu.micro05,
zhang.hpca07, madriles.isca09}. 
Although these micro helper threads are
an immensely useful concept, marshalling a very large number of micro threads
can bring practical issues:

\begin{enumerate}

\item They are largely hand crafted and carefully inserted at the
right locations, perhaps after lengthy trial-and-errors. And sometimes
the technique is intended for certain targeted applications such as
excessively memory-bound programs. A fully automated mechanism to make
these decisions may have difficulty achieving effectiveness reported in
literature~\cite{cain.ispass10}, especially when targeting a broad spectrum
of applications. 

\item A contributing factor to the manual approach is the notion of
``delinquent instructions": a few culprits created most of the performance
problems. Unfortunately, as programs get more complex and use more
ready-made code modules, problematic instructions are bound to be more
spread out~\cite{parihar.mspc13}. For instance, to account for 95\% of
last-level cache misses and branch mispredictions, an average of more than
300 problematic static instructions are involved, accounting for 10\% of all
dynamic instructions~\cite{garg.pact11}. 

\item To be effective, helper threads need to be numerous. Without
substantial hardware support (possibly on the timing critical paths), spawning
these threads, passing necessary initial values, and receiving results from
them can create significant overheads for the main thread, offsetting their
benefits~\cite{collins.isca01}.

\end{enumerate}

On the other extreme of the spectrum, an idle core in a multicore system
is used to execute a different copy of the original program on a separate
thread context~\cite{purser.micro00, barnes.micro03, zhou.pact05,
mesa_martinez.micro07, greskamp.pact07, rotenberg.ftcs99, garg.micro08,
ansari.hpca13, sundaramoorthy.asplos00, garg.pact11, kondguli.taco18,
kondguli.cal18, kondguli.asplos18}. This copy is often a
reduced version of the program (which we referred to as the skeleton) so that
it can run faster to look ahead. This style of design can be traced back to
the Decoupled Access/Execute architecture~\cite{smith.tocs84}. Unlike in DAE,
however, the leading thread in this group of designs does not affect the
architectural state and only performs look-ahead functionality. We therefore
refer to these designs as \emph{Decoupled Look-Ahead} (DLA) architectures.

DLA designs sidestep some of the practical problems facing micro helper
threads. But the key challenge becomes how to create a look-ahead thread
that is sufficiently autonomous and yet fast enough to permit deep
look-ahead. Various ways are devised to improve the look-ahead thread's
speed in order to stay ahead of the main program thread. For instance,
Slipstream~\cite{sundaramoorthy.asplos00} removes predicted dead instructions
and biased branches; Dual-core execution skips memory access instructions
that miss in the L2 cache~\cite{zhou.pact05}; Tandem uses architectural
pruning to make the hardware faster~\cite{mesa_martinez.micro07}. Garg and
Huang experimented with a more purposeful-built look-ahead thread using a
stripped-down version of the original program ~\cite{garg.micro08}.

In this past work, only a small number of ideas are discussed at a time. By
themselves these ideas have a limited benefit -- no different from ideas
for conventional microarchitectures. The limited benefit coupled with the
perceived disadvantage of doubling the resources needed can hardly make DLA
appear as a promissing solution that we believe it is. Keep in mind, the extra
thread context is an infrastructure whose cost is amortized over future ideas.
As we will show in this paper, there are many conceivable optimizations that
can lower the overhead even more while improving performance.

Other than explicitly launching a helper thread, many proposals have dealt
with reducing the chance a conventional microarchitecture is blocked~\cite{
akkary.micro03, ceze.cal04, dundas.ics97, hilton.pact09, kirman.hpca05,
carlson.isca15, martinez.micro02, mutlu.hpca03, srinivasan.asplos04,
hashemi.micro15, kadjo.micro14, hashemi.micro16, chaudhry.ieeemicro05}.
Many designs share a theme of checkpointing important state, clean up some
structures to allow further (speculative) execution. Sometimes the sole
purpose of the execution is warm-up~\cite{mutlu.hpca03}. In this latter case,
the design is more closely related to helper threading. Finally, there are
recent incarnations of the basic concept of DAE to separate the computation
part of the program from memory accesses~\cite{ham.micro15, ho.isca15}.

%% file: arch.tex
\section{Optimizations of DLA Architecture} 
\label{sec:design}

In this section, we first discuss a basic platform of DLA
(Sec.~\ref{ssec:baseline}), then discuss four optimizations in detail:
reducing look-ahead workload with prefetch offloading (Sec.~\ref{ssec:T1});
reusing value (Sec.~\ref{sssec:vr}) and control flow information
(Sec.~\ref{sssec:cfr}); and recycling the skeleton (Sec.~\ref{ssec:atb});

\subsection{Baseline DLA}
\label{ssec:baseline}

Our baseline DLA architecture is based on the one proposed in
\cite{garg.micro08}. Specifically, a \emph{skeleton} of the original program
binary is generated which includes all the control instructions and their
backward dependence chain. A subset of memory instructions is also included
in the skeleton as prefetch payloads along with their backward dependence
chain. (A more detailed discussion of the binary parsing algorithm
and parameters is left in Appendix~\ref{sec:appskeleton}.) During
execution, this skeleton forms the static code of the \emph{look-ahead thread}
\textbf{(LT)} and runs on a different core. It passes relevant information
(\eg branch outcomes) which speeds up the execution of the \emph{main thread}
\textbf{(MT)}. At first glance, it may seem wasteful to execute the same
program twice. But in reality, LT only executes a small portion of the code
(about a third in our design). Even when it does execute an instruction, the
actions involved may not be redundant. For instance, wrong path instructions
are mostly limited to the LT; off-chip accesses are only time-shifted, not
repeated; We will present quantitative description on this point later in
Sec.~\ref{sec:eval} and only note here that the energy overhead is less than
25\%.

Such an architecture requires the following support on top of a generic
multi-core architecture, ordered from least to most special-purpose:
\begin{enumerate}

\item{Containment of speculation:} LT usually involves
speculative optimizations and thus cannot be allowed to update the
architectural state. The support is simple as most of the state is already
naturally confined to the thread context. The only additional support needed
is about dirty lines in the private caches (in our study, the L1 data
cache and the L2 cache). In the look-ahead mode, they are not used
to supply coherence requests from other cores and are not written back upon
eviction, but simply discarded. In other words, when a core executes in
look-ahead mode, its private caches only obtain data uni-directionally from 
the rest system and never supplies data. It only supplies hints as follows.

\item{Communication of look-ahead results:} In a multicore architecture with
shared lower level caches, LT can already warm up the shared caches without
any additional support. However, a mechanism to explicitly pass on hints from
LT is valuable. First, we can send over the branch outcomes via the \emph{Branch
Outcome Queue} \textbf{(BOQ)}. The BOQ also acts as a natural mechanism to
detect when LT veers off the correct control flow; and to keep it from running
too far ahead. In the case an incorrect branch outcome is detected, a
\emph{reboot} is triggered by MT which re-initializes LT. In addition to
the BOQ, we can also send other information such as branch target addresses
and prefetch addresses. We use a \emph{Footnote Queue} \textbf{(FQ)} for such
less frequent but wider data. On average, one footnote is generated every 30
instructions. FQ is also used during reboot to copy the architectural
registers from MT to LT.

\item{Support for instruction masking:} Finally, we find it convenient to have
the code of LT being a subset of MT, thus allowing us to use the same program
binary and a set of bits to mask off instructions not on the skeleton. These
unwanted instructions are deleted immediately upon fetch in LT. These mask
bits can be generated either offline or online through dependence analysis
of the program binary. In this paper, we model a system where these bits
are generated offline and stored inside the program binary. At runtime, the
I-cache will combine the separately fetched mask bits and instructions.

\end{enumerate}

\begin{figure}[h]
 	\centering  
	\includegraphics[width=2.8in]{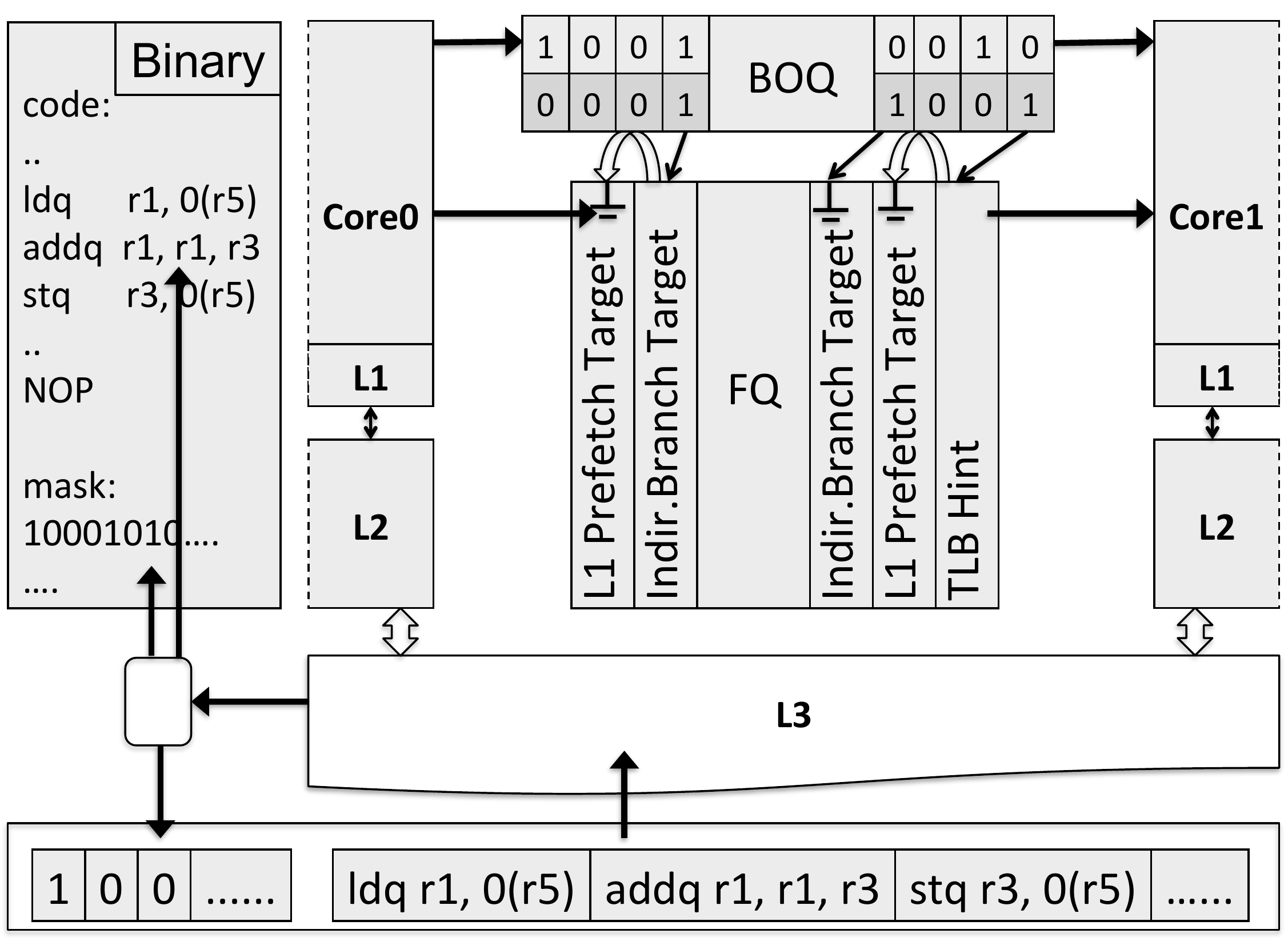}
	\caption{Architectural support for baseline DLA. \label{fig:DLA}}  
\end{figure}

\emph{Summary of operations:}
To put these elements together, we now describe the overall operation of the
system (Figure~\ref{fig:DLA}). We assume the program binary is analyzed and
augmented with mask bits offline, the system always runs in DLA mode, and that
the two threads (the real program thread and its look-ahead instance) run on
individual cores connected by the various queues discussed above. Note that
these are not intrinsic requirements to implement DLA. They describe the most
basic incarnation.

When a semantic thread is launched or context-switched in, its architectural
state is also used to initialize LT. Both threads proceed to
execute the code largely conventionally: fetching, dispatching, executing, and
committing instructions according to the content of its architectural and
microarchitectural states. They differ from conventional cores in the following
way.

For the core executing the actual program thread (MT), its fetch
unit draws branch direction predictions from the BOQ instead of its branch
predictor. If the queue is empty, we stall the fetch. The dequeued entry of
BOQ may have a footnote bit set. In that case, the control logic will dequeue
one or more entries from FQ and act according to the content type. For
example, if the entry is a branch target hint, then the content from the entry
(rather than from the core's own BTB) will be used for target prediction.

For the core executing in look-ahead mode, there are three main differences.
First, upon an instruction fetch, the logic will delete unwanted instructions.
Beyond the I-cache, however, the masks are assumed to be stored in a different
location in the program binary for backward compatibility. The masks are thus
stored separately from the instructions in the lower level caches. During an
I-cache \emph{miss}, the controller will issue two read requests to the L2
cache for the instructions (at the address $A_i$) and their masks stored in
address $A_m=f(A_i)$.\footnote{The masks can arrive asynchronously with
respect to the actual instructions. Before the actual mask bits arrive, the
system defaults to all 1s, thus including all instructions in LT. The extra
space due to masks and the separate access are all faithfully modeled in our
simulations.}

Second, LT will write hints into the queues for MT. Specifically, at commit
time, the outcome of a conditional branch (``taken" or ``not taken") will
be stored in FIFO order in the BOQ. The BOQ serves a multitude of purposes.
\ding{172} It passes a branch outcome as a prediction to MT. This ensures that
in the steady state, the majority of branch mispredictions are experienced
only in LT. \ding{173} It is a simple and effective mechanism to detect
incorrect look-ahead control flow. When a branch prediction fed by LT
turns out wrong, which is relatively rare (0.06 per kilo instructions), it
means that LT is executing down the wrong path. We will reboot LT from the
current state of MT. \ding{174} We can easily know and control the depth of
look-ahead: the number of unread entries in the BOQ equals the number of
dynamic basic blocks LT is ahead of MT. To prevent run-away prefetching, we
only need to limit the size of the BOQ (512 entries in this paper). \ding{175}
It is a convenient way to allow delayed (just-in-time) prefetching. When a
prefetch hint is generated, it can be associated with a branch entry and
released only upon the dequeuing of that BOQ entry.

Finally, in addition to the continuous branch direction hints, occasionally
LT has other hints. Whenever it encounters a miss in TLB,
L1 data cache, or BTB, it will pass the relevant address through FQ and
set the footnote bit in the most recent BOQ entry.

\subsection{R3-DLA: Overview}
\label{ssec:r3-dla}
Before we discuss the proposed
optimizations, it may be helpful to put these ideas into the context of the
vision for DLA. As we have seen, there is significant implicit parallelism
in normal programs. It is not yet clear to us what the best approach
to exploiting this parallelism is. Our current DLA design follows a path
that first targets instruction and data supply issue, basically because
we have to start \emph{somewhere}. The baseline DLA design is also just a
starting point, intuitively with many low-hanging optimization opportunities.

Indeed, as it turns out, there are a number of simple things to do to
either make the LT more efficient and/or get more utility from it. In
the first category, we can offload certain type of look-ahead code to a
finite state machine (Sec.~\ref{ssec:T1}), making the LT smaller and thus
faster. In the second category, we find that LT can provide more than just
branch prediction and addresses for prefetching. Both the intermediate
values and control flow information can be reused, for example for value
prediction (Sec.~\ref{ssec:vp}). Finally, the heuristic-driven skeleton in the
baseline DLA is clearly not optimal and can be adjusted online to other
predefined versions which, depending on the specific situation, can 
make LT more efficient and/or more effective (Sec.~\ref{ssec:atb}).

\subsection{Reduce: offloading strided prefetch}
\label{ssec:T1}

The first optimization speeds up LT by reducing its workload. The intuition
is simple. LT serves as a software-guided prefetch engine which is far more
flexible and precise than hard-wired, finite-state machine (FSM) driven
prefetchers. The cost, however, is that we need to execute a sequence of
instructions to compute the address for prefetching. For simple, strided
accesses, such a software-driven approach is an overkill. Instead, we
build a hardware FSM (which we call T1) to offload this type of prefetch.

Note that there is an important difference in the design goal between T1
and a traditional stride prefetcher. The latter needs to extract the stride
in the presence of unrelated addresses and in the absence of any certainty
that there is a strided stream to begin with. Moreover, to improve coverage,
practical prefetcher designs target variations of strided accesses. All these
are non-trivial challenges and often involves memorizing and cross-comparing a
non-trivial number of addresses. T1, on the other hand, merely carries out the
mundane task of address calculation and issuing prefetches. In other words,
compared to a traditional stride-detecting prefetcher, T1 is only a dumb FSM
carrying out simple orders. Additionally, T1 only targets one common situation
and does not try to be general-purpose.

\subsubsection{Overview of operation} The common situation targeted by T1
is a loop with one or more memory access instructions whose address gets
incremented by a (run-time) constant every iteration. In such a case, all
we need for effective prefetching is the stride ($\delta$), the prefetch
distance ($n$), and the identity of the strided access instructions. When we
encounter a strided instruction, we can take its address ($A$) and simply
prefetch $A+n\delta$. Given the identity of the strided instructions, we can
easily derive the stride and prefetch distance: The former is simply the
difference between the addresses of two consecutive instances of the same
static instruction; The latter is the average access latency divided by the
time interval between two consecutive instances. Instead of including these
necessary instructions in the skeleton to generate proper prefetches, we mark
them in MT and let T1 handle the prefetches. LT is thus smaller and faster
than otherwise. 

\subsubsection{Instruction marking} To summarize the discussion above, all the T1
hardware needs is the identities of the loop branch and the strided access
instructions. These instructions are marked with another bit (the S bit). The
S bits are generated the same time the skeleton masks are generated, however
they are a marker on the binary for MT\footnote{So, the skeleton now includes
two bits per instruction: a mask for LT and a mark for T1}. They are fetched
from the program binary by MT in the same manner as the skeleton's mask bits.
Note that the T1 hardware located in MT's core will use information from MT
to produce corresponding prefetches for the instructions marked with the S
bit. Hence, the skeleton generation process will completely ignore these
instructions and their backward dependencies while generating the skeleton for
LT.

\begin{figure}[h]
	\centering  
	\includegraphics[width=3.4in]{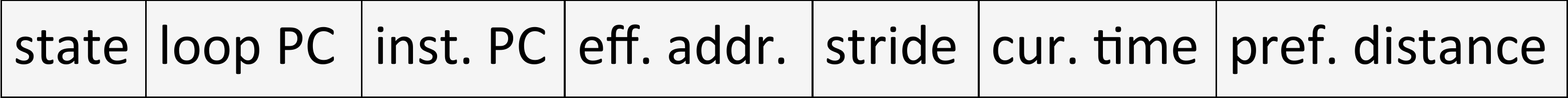}
	\caption{T1 prefetch register fields. \label{fig:T1}}  
\end{figure}

\subsubsection{Operations of FSM} With marking of S bits done, the run-time
operation is to calculate two parameters: stride and prefetching distance.
To accomplish these tasks, we use a small prefetch table. An entry of this
table is shown in Figure~\ref{fig:T1}. All entries begin in an invalid
state. Upon execution of a strided instruction, an entry is allocated in the
prefetch table. T1 starts issuing prefetches (with a fixed prefetch degree)
as soon as the first instance of a stride is calculated. Each entry in the
table gradually moves from an invalid state through transient states to the
steady state. Transient states help guard T1 against identifying incorrect
strides resulting from out-of-order execution. They also aid in calculating
appropriate prefetching distance using the information on iteration time and
average memory access latency. When prefetching distance is calculated, T1
launches multiple prefetches to ``catch up" with the prefetching distance. T1
then transitions into a steady state in which it launches one prefetch every
iteration. All entries in the table are cleared when a loop terminates.

\subsubsection{Discussion}
It may be tempting to believe that a conventional
prefetcher will competently handle all strided accesses and to question
the utility of the proposed optimization. While detailed analyses will be
presented later, a rough characterization helps. A state-of-the-art prefetcher
indeed targets almost all strided accesses. But empirically about a third of
the prefetch is still either late or evicted before access. This is why our
baseline, access-pattern-agnostic DLA system includes a non-trivial amount
of instructions in the skeleton. Every generated prefetch costs about three
instructions in LT. T1 helps eliminate 21\% of instructions from LT,
making it run faster ahead to target other incidents.

\subsection{Reuse of Value and Control Information}
\label{ssec:vp}

The software-controlled nature of the baseline DLA makes it a very flexible
branch predictor and prefetcher. The cost for such flexibility is the extra
execution. As we will see later, even with offloading discussed above, LT
still executes about half of the dynamic instructions of the program. The
next two optimizations try to increase the benefit of this work already done.
In particular, since a significant portion of the values have already been
computed, we seek to reuse them in the form of value prediction. Also, the
content of BOQ is highly accurate future control flow information and can help
improve instruction fetch for the trailing MT.

\subsubsection{Value Reuse} 
\label{sssec:vr}

A variety of techniques have been proposed to predict
values~\cite{miguel.micro16, perais.hpca14, calder.isca99,
gabbay.micro97, lipasti.asplos96, nakra.hpca99, sazeides.micro97,
sodani.micro98, zhou.ics03, thomas.pact01, tullsen.isca99, wang.micro97,
perais.micro16}. Most of these techniques rely on the history of the
values produced to predict future values. Unlike branch outcomes, a typical
value usually has non-trivial entropy and thus defy easy predictions. However,
in our system, many instructions have already been executed in LT. Empirical
observations show that over 98\% of them have the same result as their
counterpart in MT and thus lend themselves to reuse.

The basic support is similar to any value predictor: \ding{172} the predicted
value will be used to allow dependent instructions to execute early;
\ding{173} the instruction producing the value will check the outcome with the
prediction and, upon disagreement, trigger a replay. In our system, instead of
coming from a value prediction table, the predictions are read in FIFO order
fed by LT. In our design, we extend the footnote queue for this purpose: every
instruction that we decide to apply value reuse will allocate an FQ entry
containing the value and an offset indicating distance from the preceding
branch.

Again, unlike traditional value predictions, we have an abundance of highly
accurate results. Thus the key design issue for our value reuse is to
minimize the costs, which includes communication from LT to MT and the
performance loss due to incorrect values. Our approach is to limit value only
to "slow" instructions with a high confidence of successful reuse. After some
experiments, it quickly became obvious that many different heuristics can
achieve the goal. We describe one runtime version below.

At the beginning of a new loop (Sec. \ref{sssec:controller}), MT spends a
few iterations (8 in our experiments) identifying these slow instructions,
defined as having dispatch-to-execute latency of at least 20 cycles.
Their PCs will be recorded in a bloom filter (let us call that \emph{Slow
Instruction Filter} or \textbf{SIF}). LT checks this table at commit stage and
if the instruction is there, allocates a \textit{value reuse} entry in FQ. The
SIF is cleared upon entering a new loop.

Our confidence mechanism is simplistic: when a value prediction turns out
incorrect, the entry of that static instruction is deleted from the SIF and
LT will no longer provide a prediction for that instruction. However, we
observe that this is infrequent (less than twice per million instructions).
Finally, in the typical implementations where memory consistency is
ensured with proper replays~\cite{yeager.ieeemicro96, tendler.ibmjrd02}, a
value-predicted load is considered executed at the time of validation for the
purpose of replay tracking~\cite{martin.micro01}.

\begin{figure}[htb]
	\centering  
	\includegraphics[width=2.8in]{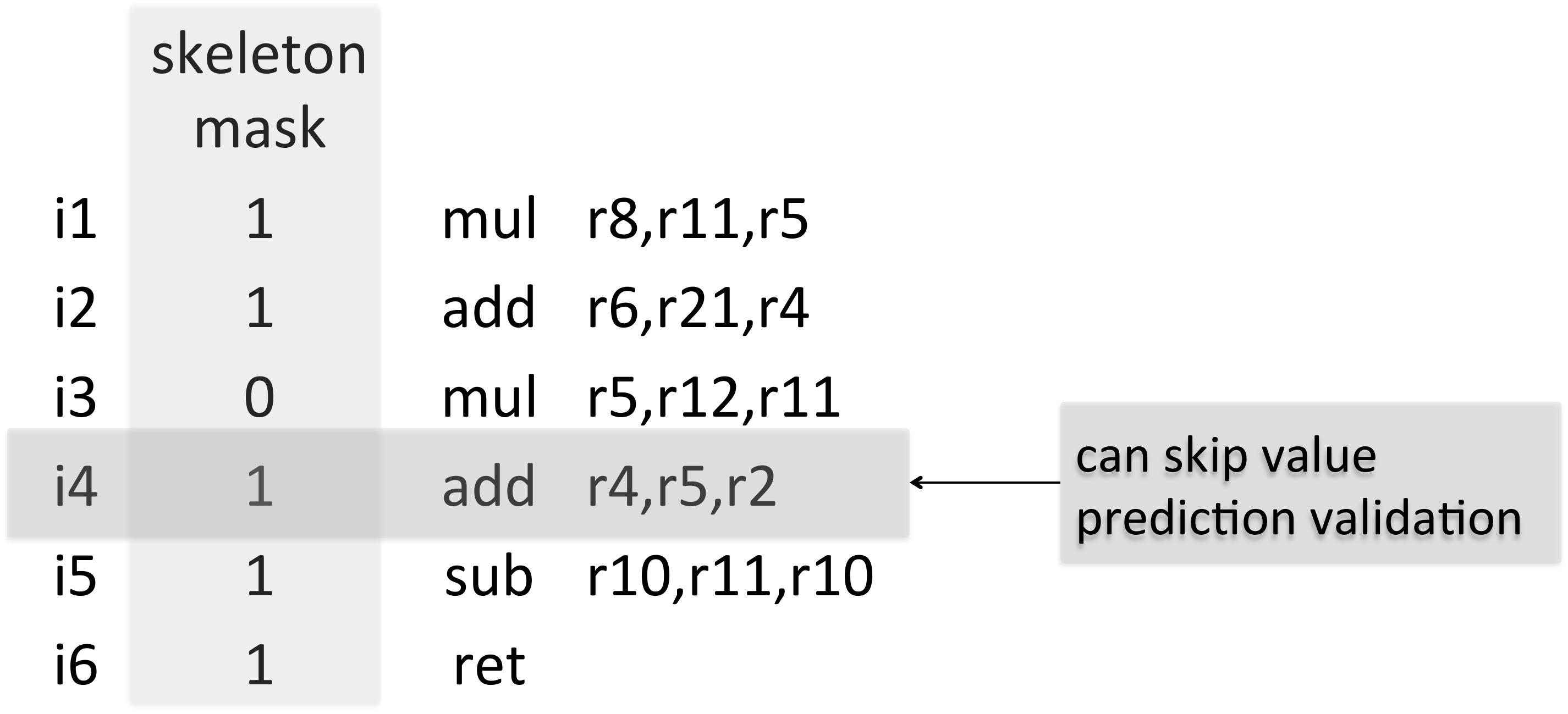}
	\caption{Example of skipping value prediction validation.\label{fig:vp}}
\end{figure}

There is one small optimization to this basic support. In some cases, we do
not need to validate all predicted values. We can skip those ALU instructions
that only depend on other instructions that have produced a predicted value.
Figure~\ref{fig:vp} shows an example. $i_4$ sources from $i_1$ and $i_2$, both
of which produce a value prediction. When we see this case, we can directly
use $i_4$'s value prediction as the outcome and there is no need to execute
$i4$ in MT. This is because in our system, we have no speculative optimization
that can corrupt functional units in LT. So, if both values are correct, then
$i_4$'s result is correct (barring hardware reliability issues). If either
value is incorrect, eventually, there will be a value misprediction recovery
upstream. $i_5$, on the other hand, cannot skip validation as it depends on
a value that is not predicted. This optimization will save about 11\% of
validations.

We implement this optimization with a simple score-boarding logic at the
decode stage in the MT core. When an ALU instruction $i$ produces a value
prediction, we mark its destination register as validated. Other instructions
(e.g., loads, or instructions not producing a value prediction) will clear the
marking for its destination register. If an instruction has a value prediction
and its source registers are all marked validated, it will not be executed for
validation. 

Finally, once the value prediction framework exists, we can add some
critical-path instructions back to the skeleton. Clearly the trade-off is
faster execution of MT at the expense of slowdown of LT. In general, whether
adding an instruction to the skeleton speeds up the whole system or not
depends on the balance of the duo. It opens up a general optimization problem
of choosing the right skeleton that maximizes system performance. In this
paper, we only follow a simple heuristic to find candidates: they have a long
dispatch-to-execute latency (more than 20 cycles on average) and have more
than one dependent instruction. The skeleton construction algorithm will
include the necessary backward dependence chain.

\subsubsection{Control flow information reuse}
\label{sssec:cfr}

Instruction fetch can sometimes be a source of pipeline bubbles. In DLA,
the presence of future control flow information allows us to ameliorate
this problem to some extent for MT. For instance, a trace
cache~\cite{rotenberg.micro96} can increase the number of fetched instructions
per cache access. Having a highly accurate stream of branch predictions from
the BOQ is a significant advantage to leverage when using a trace cache.
However, trace cache is an expensive form of instruction caching. In this
paper, we opt for a simpler approach that is in fact more effective in our
setup.

The basic idea is to reduce idling for the instruction fetch unit by
allowing it to continue even if the decode stalls. In other words, we want to
decouple the fetch unit from the rest of the pipeline -- or in the case where
this has already been done to the baseline architecture, increasing the degree
of decoupling with a bigger buffer. The key point to emphasize is that the
BOQ offers a much higher degree of branch prediction accuracy. Without this
accuracy, fetching too much down the predicted control flow is unlikely to
pay off, and indeed can even backfire and slow the whole processor down. In
fact, in a conventional architecture, sometimes a more constrained fetch unit
is beneficial as it slows down the pollution created by the relatively common
wrong-path instructions. In other words, the benefit of having a fetch buffer
is clear for DLA, but not necessarily so for a conventional architecture. We
will show this in the experimental analysis later in Sec.~\ref{ssec:detailed}.

To understand the effect a bit more, we will perform a simplified,
first-principle analysis here to show how much a fetch buffer can improve the
fetch performance and whether having a trace cache will materially improve
the effect. We can measure the performance of a fetch unit by how many
\emph{fetch bubbles} it inserts into the pipeline downstream, \ie how many
more instructions the next stage (decode) can absorb but the fetch unit fails
to deliver. This can be calculated from a simplified probabilistic 
analysis~\footnote{A more detailed mathematical reasoning behind this analysis 
is presented in Appendix~\ref{sec:appfb}.}.

Figure~\ref{fig:wabi} shows the effect of the fetch buffer under the
simplified probability model. Figure~\ref{fig:wabi}-a shows the probability
distribution of queue length. We see that a key outcome of having a larger
capacity is the reduction of probability of an empty queue. This means that a
longer buffer allows the fetch unit to use otherwise idle cycles to fill the
queue fuller to reduce the impact of, for example, an instruction cache miss.
Trace cache, on the other hand, only marginally increases the filling speed.
Without a longer buffer, its effect is very limited. With a longer buffer, it
becomes essentially unnecessary.

\begin{figure}[htb]
 	\centering  
	\subfigure[]{\includegraphics[width=1.5in]{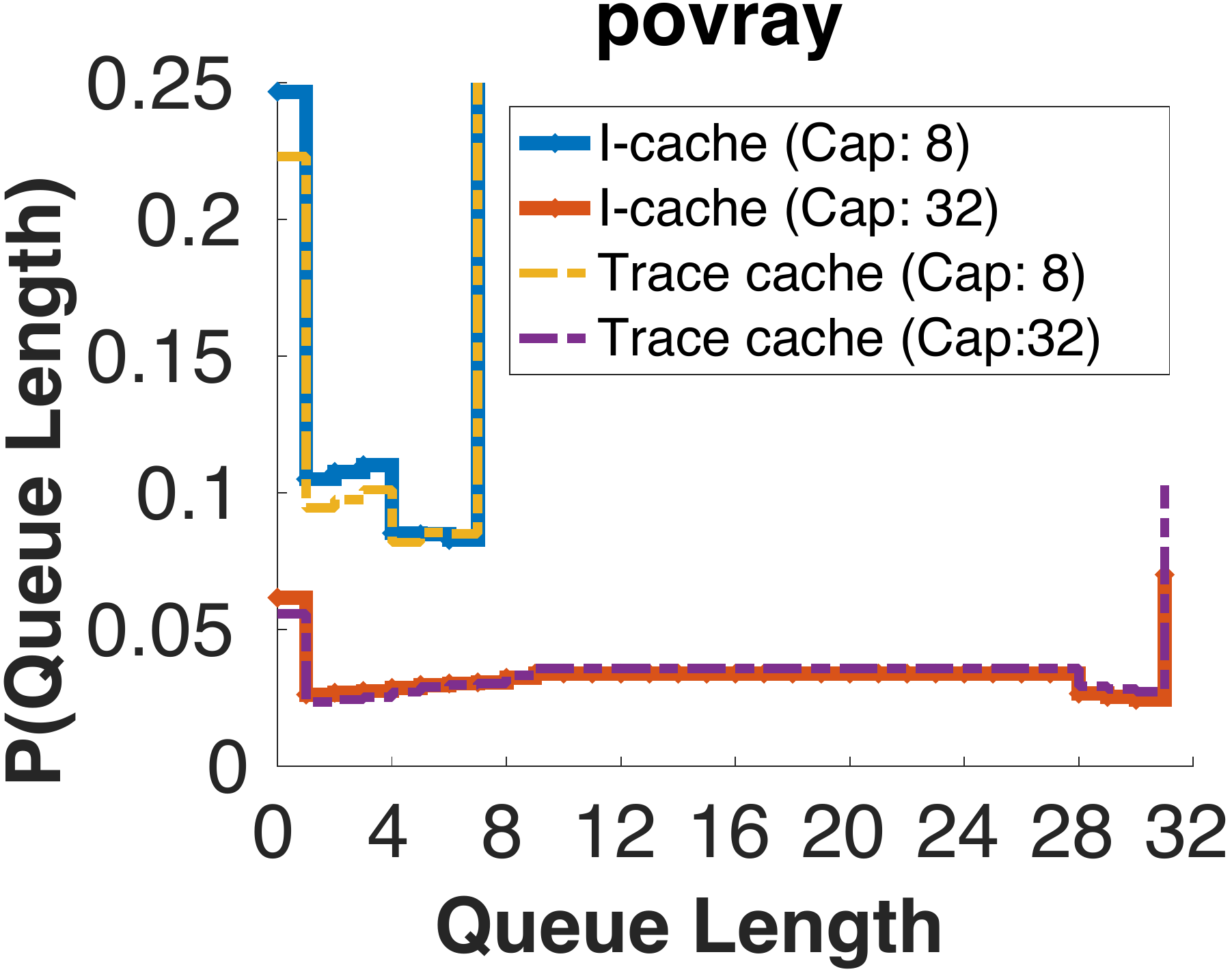}}
	\subfigure[]{\includegraphics[width=1.5in]{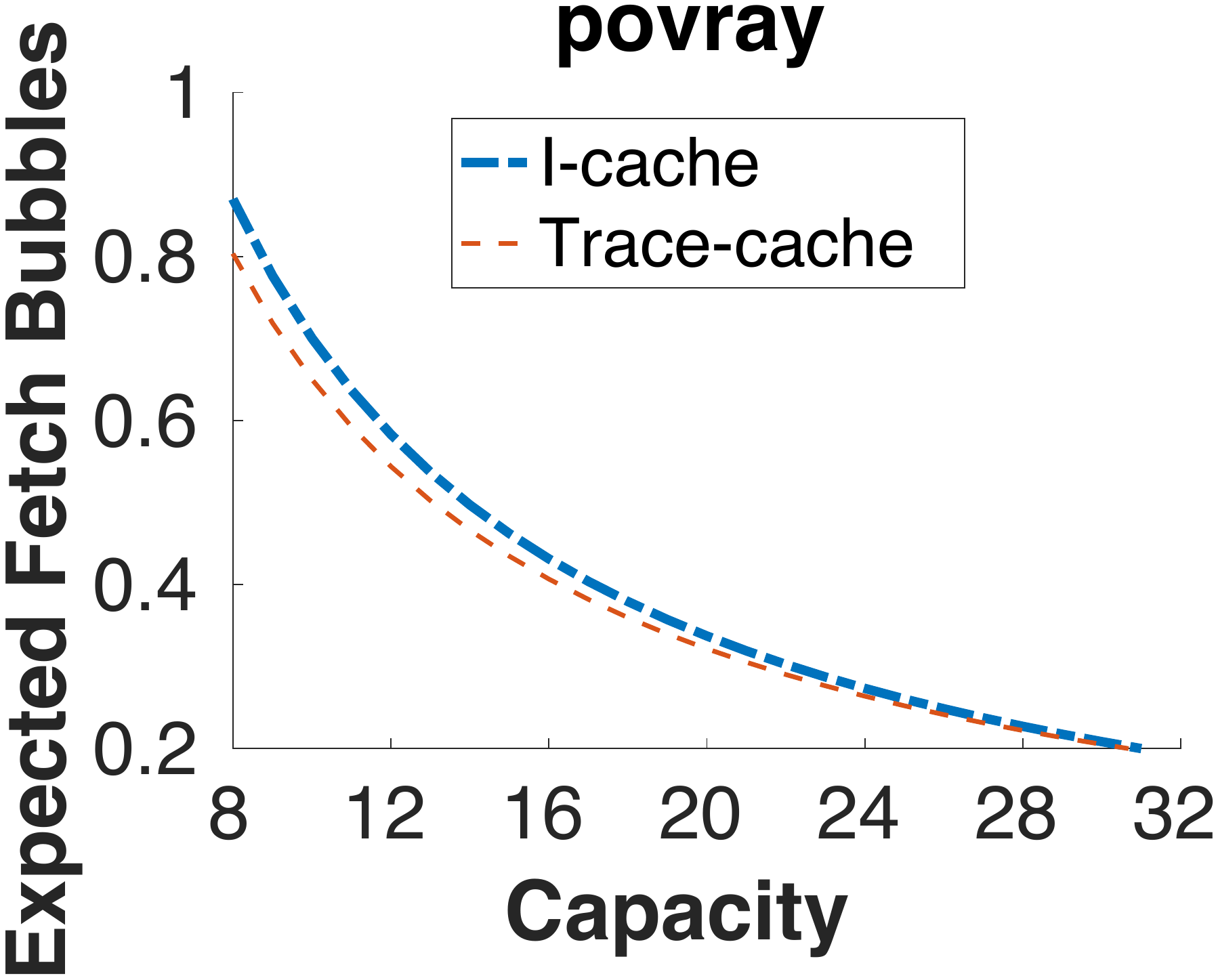}}
	\caption{(a) The estimated probability distribution of queue length
	with a queue capacity of 8 and 32 entries under both I-cache and trace
	cache. (b) The expectation of fetch bubbles as the queue capacity varies.
	In this analysis, the probability distributions of fetch demand and
	supply are empirically obtained from one application (\texttt{povray}).
	\label{fig:wabi}} 
\end{figure}

In fact, Figure~\ref{fig:wabi} shows the application with the most pronounced
difference between the two caches. In other applications, the difference is
even smaller. From Figure~\ref{fig:wabi}-b we see that with a small increase
in fetch queue capacity, the number of expected bubble drops from more than 1
per fetch to less than half.

Note that the benefit of increasing fetch buffer size is closely tied to the
branch prediction accuracy. This is because the analysis above assumes that
every instruction in the queue is on the right path and therefore useful. This
is a reasonable assumption in our case as the effective branch prediction
accuracy from the BOQ is above 99.9\%. In a conventional architecture, the
elevated misprediction probabilities make this analysis inaccurate. In fact,
sometimes a more constrained fetch unit is beneficial as it slows down the
pollution created by wrong-path instructions. In other words, the benefit
of having a fetch buffer is clear for DLA, but not necessarily so for a
conventional architecture. We will show this in the experimental analysis
later.

\subsection{Re-cycling the Skeleton}
\label{ssec:atb}

In DLA, the skeleton is constructed using simple heuristics. Given this basic
skeleton, if we add one more memory instruction (and its backward dependence
chain), LT is likely to run a bit slower but potentially
helping MT avoid more misses. Depending on which thread tends
to be the bottleneck, this small change may increase or decrease system
performance. We can see that there is a vast number of possible variations
and the basic skeleton is unlikely to be optimal. The question is, therefore,
are there significantly better options than our default? If so, how can we
systematically and efficiently arrive at such options?

These are all questions beyond the scope of this paper. Nevertheless, we do
know that simple tunings can effectively improve the performance. The general
approach is to create a few versions of skeleton and cycle through them to
find out the best empirically.\footnote{This process may repeat a few times to
average out noise. Admittedly, the analogy with recycling in the normal sense
is tenuous.}

\subsubsection{Versions of skeleton}
 
The most basic version of the skeleton
includes all branches and their backward dependence chains and is produced
with a binary parser. From this starting point, we may add or subtract
instructions using a few broad heuristics coupled with static-time profiling.
In our experiments we collect these statistics by executing the programs
with training inputs and use them to build skeletons that are used during the
actual run.
We experimented with five options:
\begin{itemize} 

\item{L2 prefetch targets:} Instructions that account for significant 
portions of L2 misses can be added to the skeleton;

\item{L1 prefetch targets:} Instructions that account for 
significant portions of L1 misses can be added to the skeleton;

\item{Value reuse targets:} Instructions that have a long dispatch to execute
latency can be added to the skeleton;

\item{T1 targets:} Memory instructions that are handled by T1 are by default
removed from the skeleton. However, they may be added back (as they might warm
up cache for LT).

\item{Biased branches:} Conditional branches with a bias over a threshold
can be converted to unconditional branches in the skeleton.

\end{itemize}

\begin{figure}[t]
    \centering
    \includegraphics[width=3.4in]{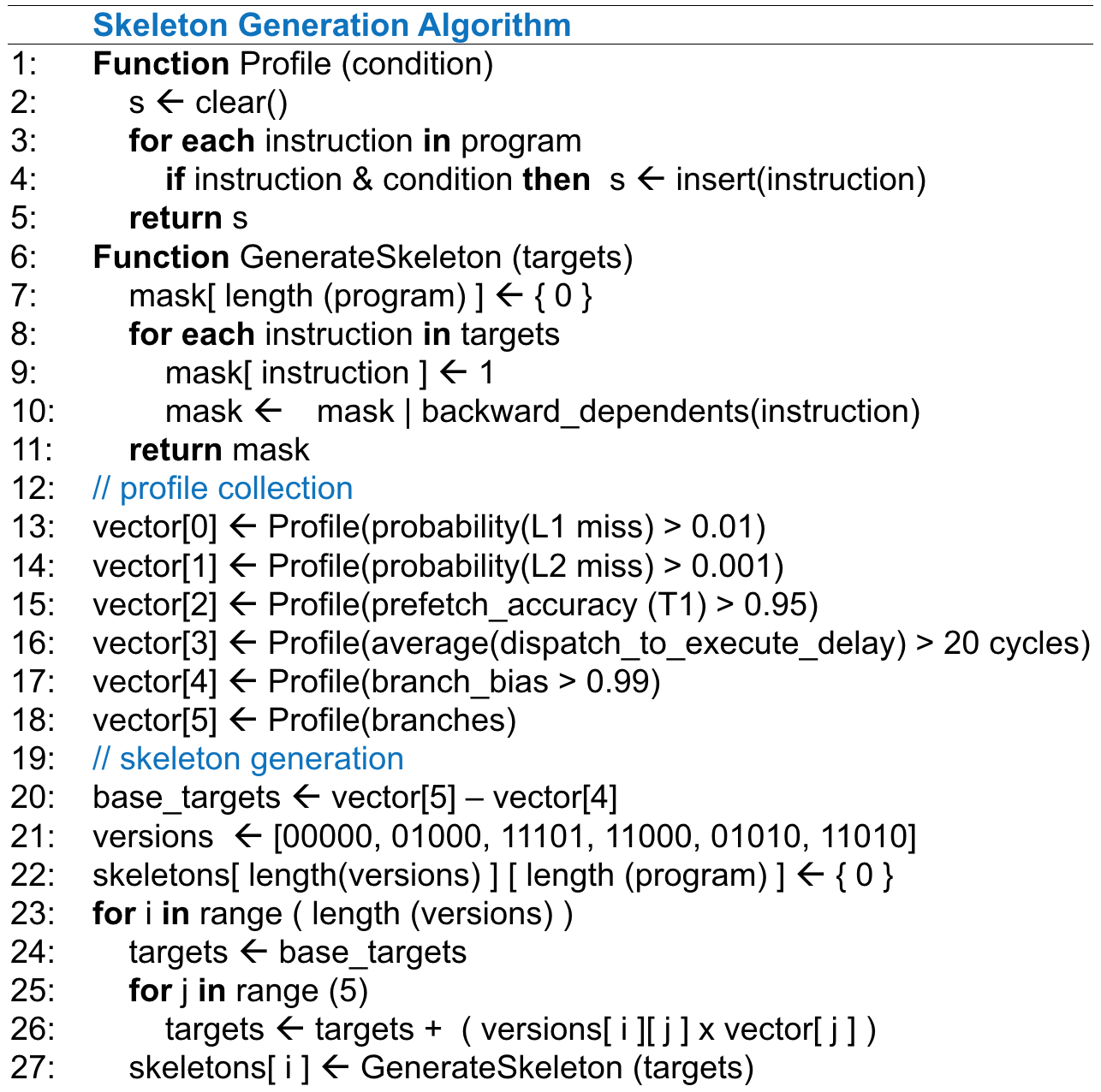}
	\caption{A pseudo code outlining the steps involved in generating different
skeletons used by the recycle optimization. A few seed vectors are constructed
using profiling information obtained from the program binary and training runs.
A skeleton is generated from a seed vector by including backward dependencies of
each seed present in it. Multiple combinations of these seed vectors can
therefore produce multiple skeletons. The recycling optimization we used in our
evaluations uses six of these skeletons (Line 21 in the figure).\label{fig:sktgen}} 
\end{figure}

These independent options naturally lead to many different combinations.
Our empirical observation shows that a very small number of combinations
need to be searched to obtain noticeable benefits. We evaluate a design that
cycles through six versions of skeleton empirically observed to be most
often helpful. Changes to the number of options, the number of versions, or
the thresholds used in identifying target instructions will likely affect
the exact outcome. The key point to note is that this is not an effort
to find the optimal points in the design space, but simply an attempt to
pick a few different points so that we can avoid poor design points due
to simplistic heuristics. Also note that the skeleton generation process
(Fig.~\ref{fig:sktgen}) is an offline, automated process just as in the
baseline, except it produces more than one skeleton.

\subsubsection{Controller} 
\label{sssec:controller}

With a number of skeleton options, the goal of the
controller is to find the skeleton version that maximizes the benefit. To do
this, we divide the execution into repeating code units, or loosely speaking,
loops. For each loop, we cycle through different versions to find out the best.

\begin{figure}[htb]
    \centering
    \includegraphics[width=3.4in]{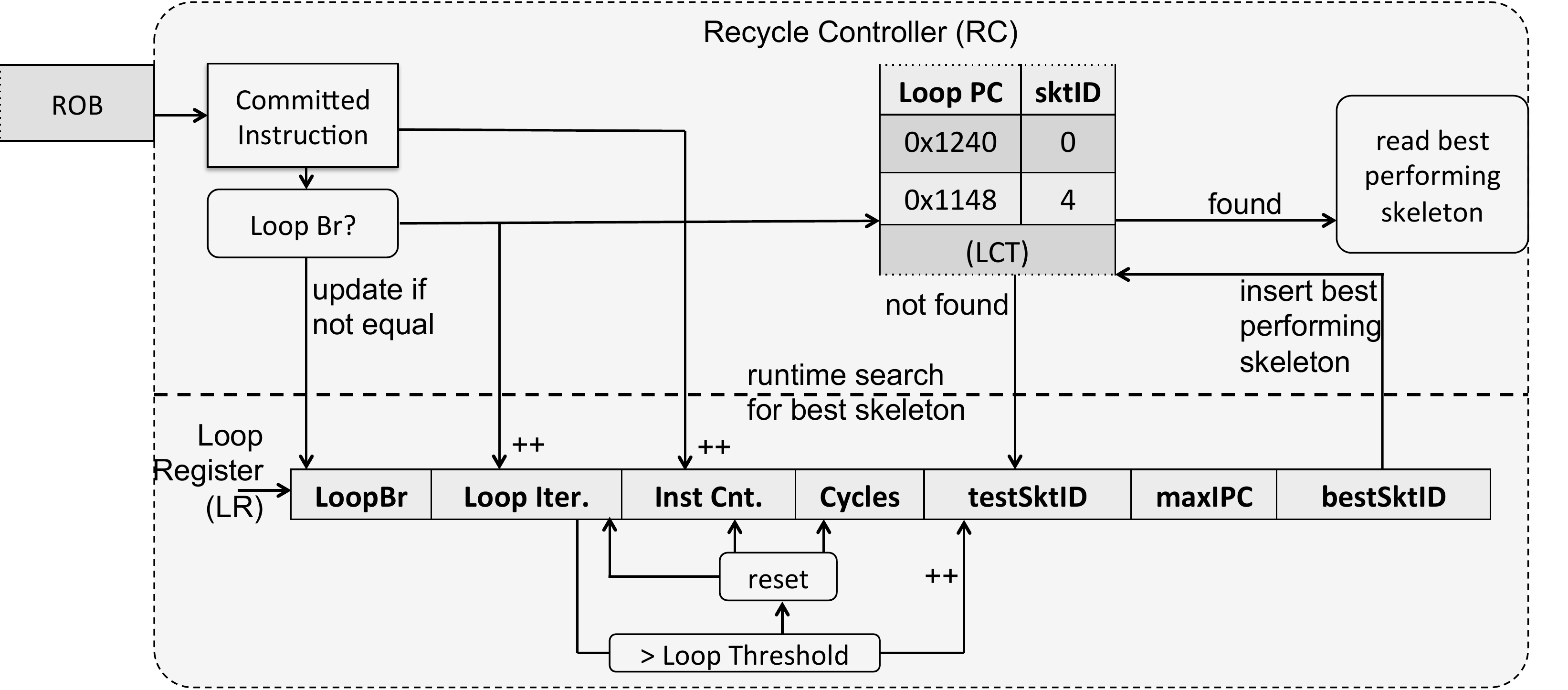}
	\caption{Skeleton recycling flow chart. As a loop branch retires, the Loop
Config Table (LCT) is queried for the skeleton that is optimum for the current
loop. If none of the entries in LCT match the loop branch, different fields in
the Loop Register (LR) are used to cycles through each of the available
skeletons for a few iterations of the loop and identify the optimum skeleton for
the loop. The LCT is updated when an optimum skeleton is found and that skeleton
is used by lead thread until a new loop branch retires from ROB.\label{fig:loopmarking}}  
\end{figure}

\begin{figure*}[b]
 	\centering  
	\includegraphics[width=6.8in]{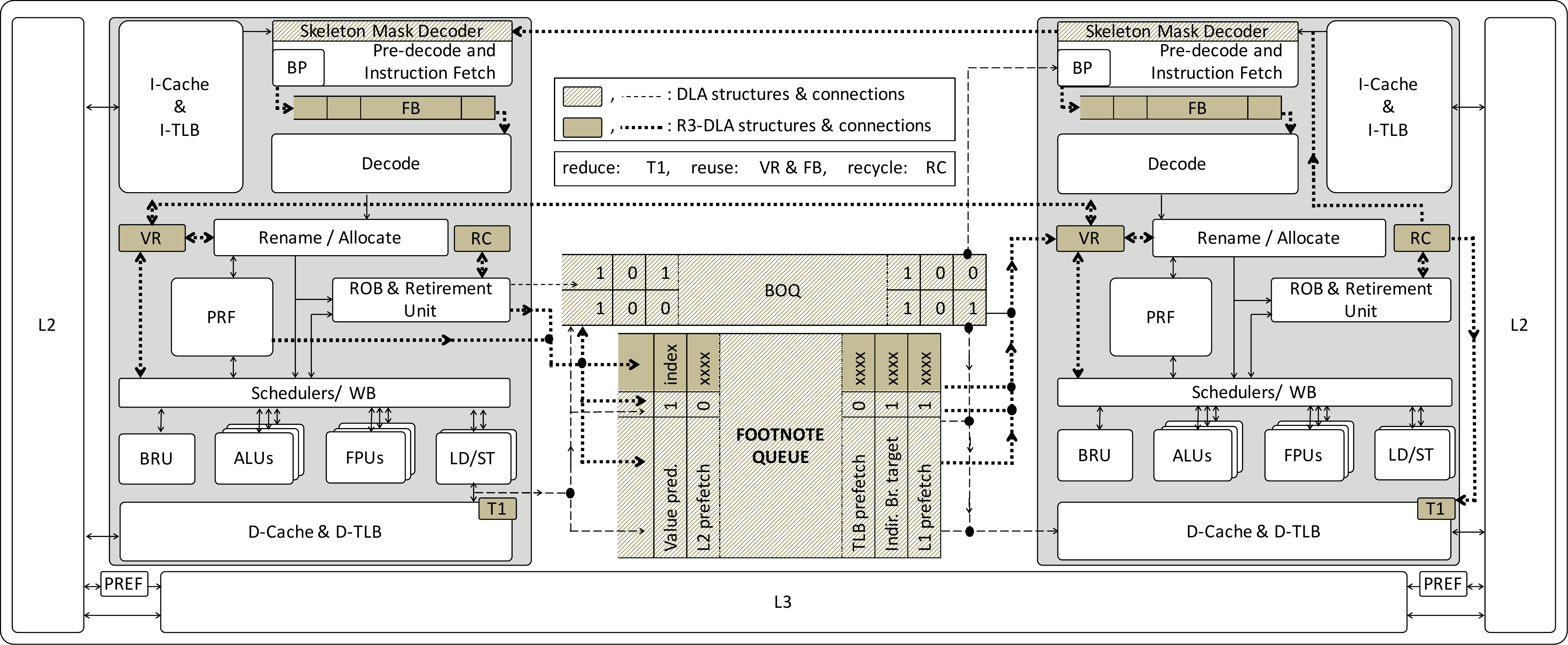}
	\caption{Architectural support for R3-DLA. The gray colored rectangle on the
left represents the lead core and the one on the right represents the main core.
Structures included by DLA are patterned and the connections are indicated with
the dashed lines. Structures included by R3-DLA are shaded and the connections
are indicated with the dotted lines. \label{fig:R3DLA}}  
\end{figure*}

To identify the current loop, we capture the backward ``loop branch"
(Figure~\ref{fig:loopmarking}). Two consecutive instances of the loop branch
without an interleaving instance of another loop branch marks the two ends of
an iteration. Note that units need to be of sufficiently coarse granularity,
otherwise we can neither accurately measure execution statistics nor
profitably adjust the system configuration. So, a unit of execution is one or
more iterations lasting, say, at least 10,000 instructions.

Note that recursive functions can represent a significant portion of the
execution time without having a detectable loop. To deal with these cases,
we treat certain function call instructions as if they are loop branches. In
such a case, an ``iteration" may not have the same meaning as we are used to.
But it is still a valid strategy to observe the behavior of a unit of multiple
iterations to predict that of a future unit.

During execution, the controller will run each loop for enough iterations
under a particular skeleton. This will allow an accurate measurement of the
speed (instructions committed per unit time) of that loop under that skeleton.
After cycling through all skeletons, the controller selects the skeleton
showing the highest speed and use it for the steady state. The identity of the
loop (PC of the loop branch) and the corresponding best configuration is then
stored in a Loop-Config Table (LCT as shown in Figure~\ref{fig:loopmarking}).
If a loop PC is found in the LCT, the controller selects the corresponding
configuration.

Note that all the steps involved in re-cycling skeleton can be done either
on-line as the application runs, or off-line using training runs. For the
simple recycling discussed in this paper, we believe the offline approach is
more advisable as we need no architectural support (other than performance
counters). However, an online recycling support (like the one we discussed)
may be a better alternative in a more dynamic environment. We will compare the
effect later in Sec.~\ref{ssec:detailed}.

\subsection{Recap}

To sum up, a basic DLA design uses one core to execute a look-ahead thread and
passes information through two queues (BOQ and FQ) to help accelerate MT.
On top of this basic design, we propose to add a number of supporting
elements (Figure~\ref{fig:R3DLA}) to accelerate either LT
or MT: \begin{itemize}

\item{\bf T1:} A prefetching FSM to offload prefetching of loop-based strided
accesses;

\item{\bf Value reuse:} logic to pass register values from LT
through FQ and used as predictions in the front-end of MT;

\item{\bf Fetch buffer:} using an extended buffer to fetch instructions down the
path predicted in the BOQ;

\item{\bf Re-cycle controller} (hardware support optional): cycles through a
number of skeleton mask bits to pick the best performing configuration.

\end{itemize} With these elements, the R3-DLA becomes significantly more
effective as we will show next. More importantly, these optimizations are
merely examples of what can be done to make the DLA models more effective. We
believe that there are plenty of opportunities in DLA to keep on extracting
more implicit parallelism.

%% file: eval.tex
\section{Experimental Analysis}
\label{sec:eval}

In this section, we perform experimental analyses of the proposed design.
After detailing the simulation setup (Sec.~\ref{ssec:setup}) we first show
bottom-line results of a complete system (Sec.~\ref{ssec:sysperf})
and then provide more detailed analyses to gain insight of the individual
design decisions (Sec.~\ref{ssec:detailed}).

\input{env}

\subsection{Overall Benefits}
\label{ssec:sysperf}

It is worth noting that in this paper, we assume DLA is only used as a
turbo-boosting technique -- when there is an idle core/thread. We assume
otherwise exploiting explicit parallelism yields better results.\footnote{Note
that this is not always true anymore. And as more ideas are developed, we
may reach a point where the decision whether to use resource for one type of
parallelism or the other becomes a non-trivial one.} 

\begin{figure*}[t]
 	\centering  
	\subfigure[]{\includegraphics[width=1.68\columnwidth]{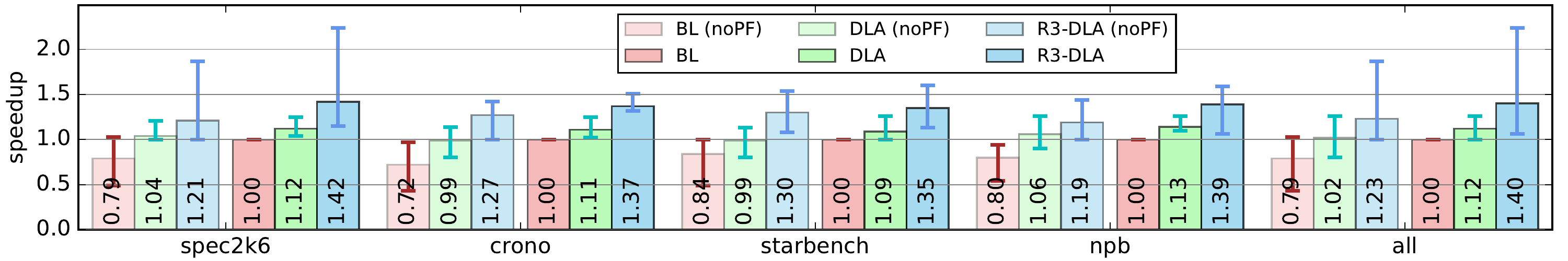}}
	\subfigure[]{\includegraphics[width=0.32\columnwidth]{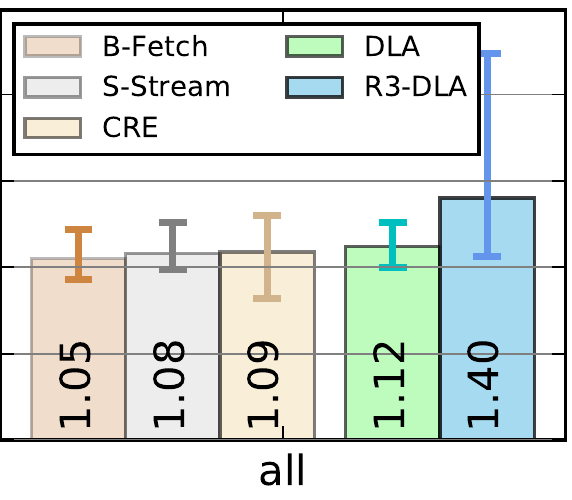}}
	\caption{Performance gain over an aggressive baseline with BOP at
	L2. NoPF shows the normalized performance of a baseline configuration with
	no prefetcher.  \label{fig:perf}} 
\end{figure*}

\subsubsection{Performance} We first measure the overall performance of R3-DLA and
compare it to that of the underlying microarchitecture and that enhanced by
our baseline DLA. We show these three configurations both without a hardware
prefetcher (left group of 3 bars In Figure~\ref{fig:perf}-a) and with BOP
prefetcher (right group of 3 bars). All performance results are normalized
to the microarchitecture with BOP, which represents the best we can do today
without using DLA techniques.

For clarity, we summarize the results of an entire benchmark suite into a
single bar showing the geometric mean of the whole suite and an I-beam showing
the range of values. This figure contains a lot of information that can be
organized into a number of observations: \begin{enumerate}

\item R3-DLA provides high performance compared to the underlying
microarchitecture with an advanced prefetcher. The speedup ranges from
1.06x to 2.24x with a geometric mean of 1.4x. While the average
performance gain is significant, there is also a wide range of result. DLA is
most likely to be selectively applied when the benefit is large. For instance,
for the top half of the applications, the geometric mean speedup would be
1.51x.

\item R3-DLA is also significantly faster than more basic DLA designs.
On average, R3-DLA outperforms the baseline DLA by about 1.25x.
This shows that the proposed optimizations are effective.
Figure~\ref{fig:perf}-b briefly compares the overall performance among a set of
related approaches: B-fetch~\cite{kadjo.micro14},
SlipStream~\cite{purser.micro00} and CRE~\cite{hashemi.micro16}.

\item DLA is a fully-flexible prefetcher and thus has overlapping targets
with a standalone prefetcher such as BOP. When used with R3-DLA, BOP can still
help as it frees up DLA's attention to better handle the remaining targets,
making the system a bit more efficient. However, the ``collaboration" between
the two mechanisms is unplanned for and the benefit is somewhat limited: while
BOP can improve the baseline architecture by 1.27x, its effect on an R3-DLA
system is only 1.13x. We conjecture that a more conscious collaboration between
a standalone prefetcher and DLA will be more effective.

\end{enumerate}

\subsubsection{Efficiency} One common concern of DLA architectures is the energy
cost. While it is tempting to assume the energy cost (or at least power)
doubles in DLA due to executing the program twice, it would be a significant
overestimation even for the baseline DLA design, not to mention R3-DLA, which
further lowers the overhead. 

First and foremost, LT is a much lighter thread, with an average length of
only 36\% that of MT\footnote{Some instructions (about 7\%) are prefetch
instructions and do not enter the commit stage. The committed instructions
in LT, therefore, amount to about 29\% on average.}. Second, not all LT
activities are overheads. Some are time-shifted activities (\eg most memory
accesses). Others help MT avoid almost all wrong-path instructions. Finally,
faster execution lowers fixed energy costs. Note that LT-to-MT communication
is insignificant (averaging 2.2 bits per instruction) and is faithfully
modeled. To see this in a bit more detail, in Table~\ref{table:energy}, we
show the amount of activities, the resulting dynamic and static power in both
LT and MT, all normalized to the baseline microarchitecture. We see that LT
expends much less dynamic energy or power than baseline. Also, despite running
much faster than baseline, MT's power is comparable to the latter since it
significantly reduces waste.

\begin{table}[htb]
\scriptsize
\centering
\begin{center}
  \begin{tabular}{|p{0.36in}|p{0.12in}|p{0.14in}|p{0.14in}|p{0.16in}|p{0.28in}|p{0.2in}|p{0.2in}|p{0.2in}|}
    \hline 
	 &  & D & X & C & Dyn. Energy& Dyn. Power& Static Power & Power\\
    \hline
	\multirow{2}{*}{DLA}	   & LT & 49\% & 48\% & 48\% & 48\% & 54\% & 94\% & 71\% \\
							   & MT & 77\% & 86\% & 100\% & 88\% & 96\% & 99\% & 97\% \\   
    \hline 
	\multirow{2}{*}{R3-DLA}    & LT & 35\% & 29\% & 29\% & 30\% & 42\% & 93\% & 64\% \\ 
							   & MT & 77\% & 82\% & 100\%  & 80\% & 110\% & 95\% & 103\% \\ 
    \hline 
  \end{tabular}
\end{center}
	\caption{Average of activities (in Decode, eXecution, and Commit stages),
	energy, and power for both threads in DLA and R3-DLA all normalized to
	baseline. Note that for every instruction committed in the baseline
	processor, 1.16 are executed and 1.25 decoded.} \label{table:energy} 
\end{table}

\begin{figure}[h]
	\centering   
	\subfigure[]{\includegraphics[width=0.48\columnwidth]{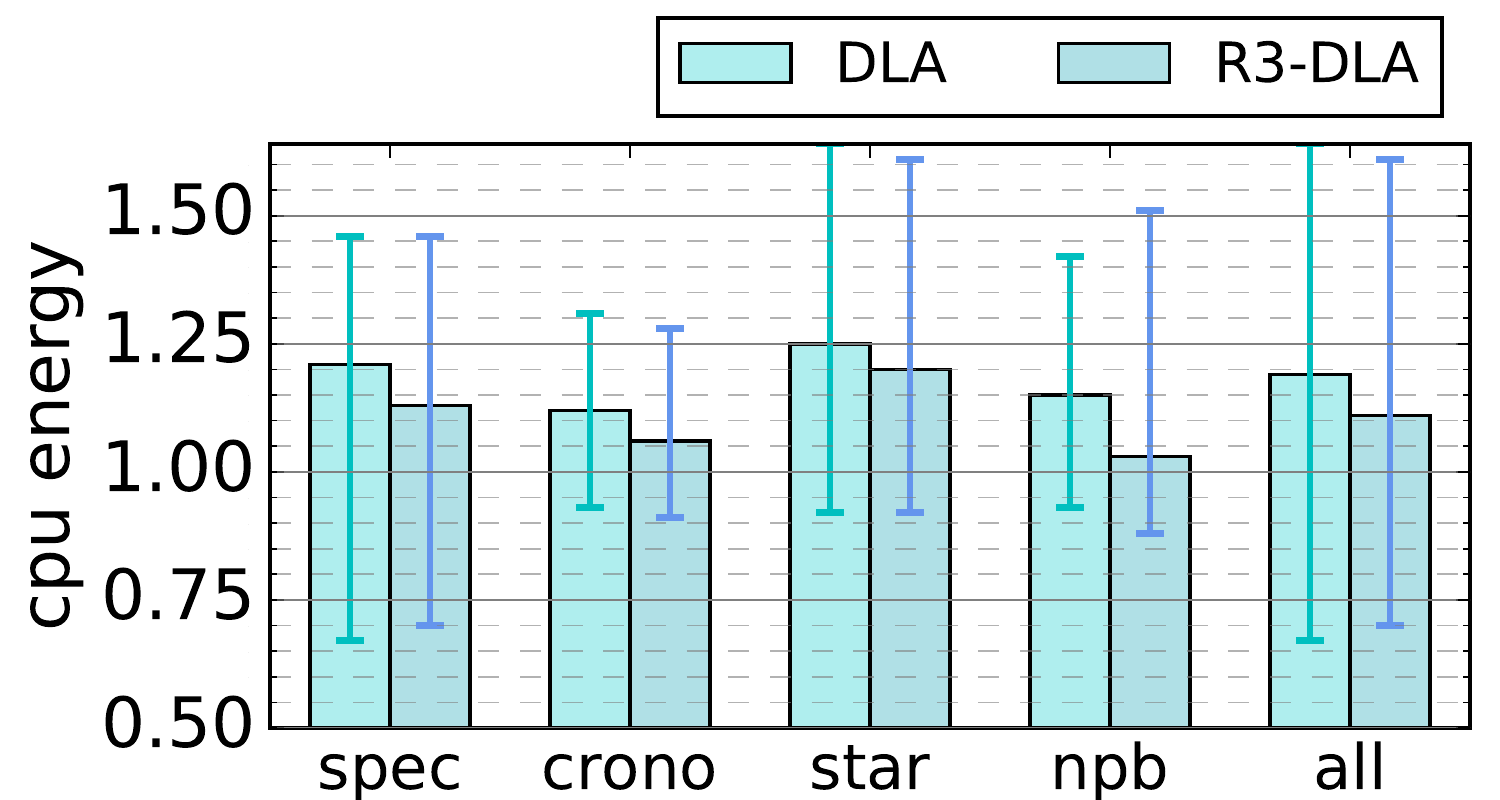}}
	\subfigure[]{\includegraphics[width=0.48\columnwidth]{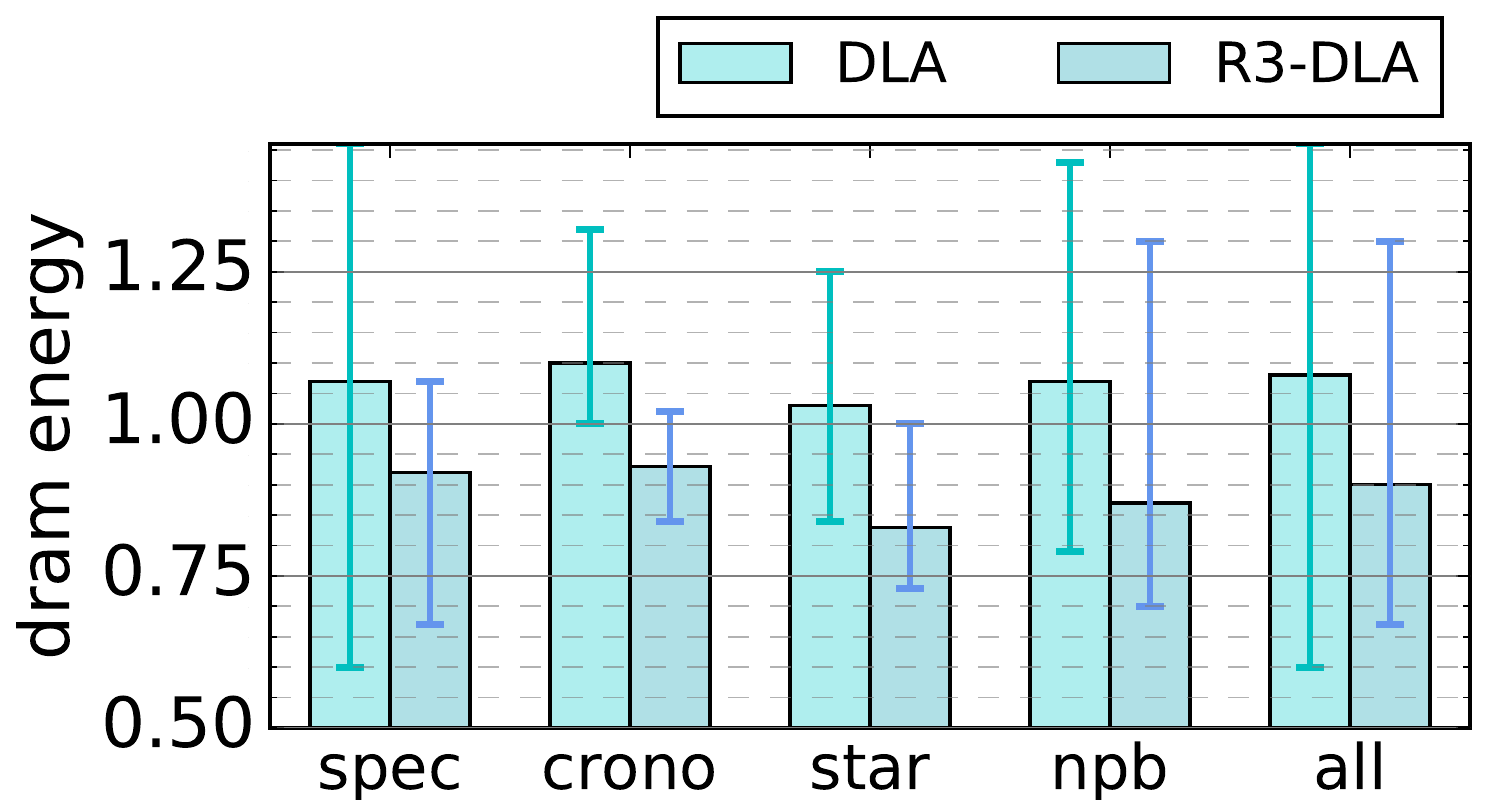}}
	\caption{Comparison of energy normalized to baseline spent in (a) cpu (b)
	dram by DLA and R3-DLA.\label{fig:energy}} 
\end{figure}

\begin{figure*}[b]
	\centering   
	\includegraphics[width=2\columnwidth]{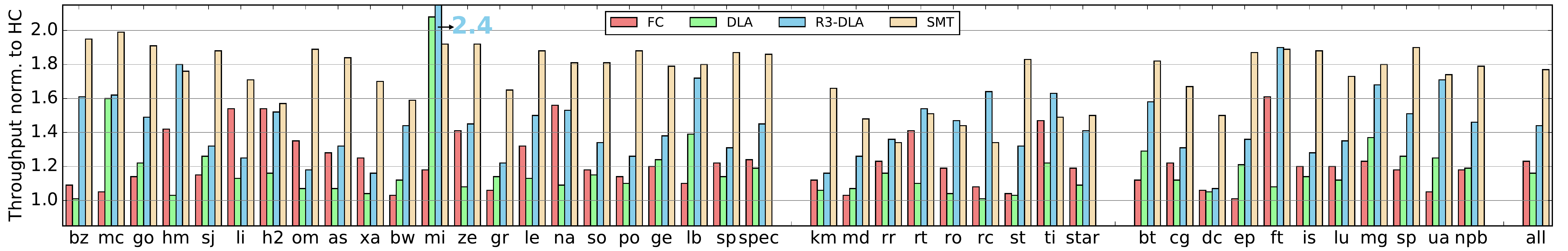}
	\caption{Comparison of normalized throughput obtained by R3-DLA using a single wider
	core over a half-core (HC). \label{fig:smt}} 
\end{figure*}

Combining these factors together, our energy estimates
(Figure~\ref{fig:energy}) suggest that the average normalized energy for
R3-DLA is 1.11x for the processor and 0.9x for memory (all geometric means).
There is significant variation among individual benchmarks (with arithmetic
mean 1.19x and 0.92 respectively.) In terms of energy delay
product, DLA is 6\% worse than baseline while R3-DLA is 19\% better on
average.

\subsubsection{Application in SMT cores}
Finally, with increased efficiency and sophistication, R3-DLA opens up the DLA
architecture to more usage scenarios. Here we show one example of SMT cores.
These cores are a good compromise between pursuing single-thread performance
and throughput. When enabling DLA on an SMT core, we use one thread to perform
the look-ahead and another to run the main program thread. Our wide SMT core
is loosely modeled after IBM POWER9 SMT8 where the core can also function
as two independent, narrower cores -- which we call half-core here. The full
core has a fetch/decode/issue/commit width of 16/12/16/16 with 512 ROB entries.
We also model the branch predictor presented in Table~\ref{table:config} for
this core. The cache hierarchy, prefetcher and memory configurations for this
core are modeled with the same parameters as presented in
Table~\ref{table:config}.

Figure~\ref{fig:smt} shows the performance results. We compare four
different usage scenarios: \ding{172} FC, which uses the entire wide core for
single-thread execution; \ding{173} DLA, which uses the SMT core to always run
two threads (the look-ahead and the main thread) on two half-cores, \ding{174}
R3-DLA; and finally, for reference,\ding{175} SMT, where we show the
throughput improvement of using the wide core to run two copies of the same
benchmark. The only modification to R3-DLA here is that in the recycling
technique, we also allow an empty skeleton, which allows all the resource to be
used for the main thread. We normalize the result to that of the half-core.

As we can see, on average, a wider core is indeed not always effective in
improving single-thread performance: though speedup can be as high as 1.61,
the global average is only 1.23. DLA can sometimes be a much more effective
technique, reaching a speedup of 2.08. But on average, it is not as effective as
a wider core. R3-DLA is significantly better than both (with a speedup of
1.44) and represents an important step towards supporting the goal of high
single-thread performance. 

\subsection{Detailed Analysis}
\label{ssec:detailed}

We now look at the contribution of each individual element in the design and
some aspects of their interaction.

\subsubsection{Offloading strided prefetch}
\label{sssec:T1_eval}

\begin{table}[htb]
\scriptsize
\centering
\begin{center}
  \begin{tabular}{|p{0.22in}|p{0.18in}|p{0.21in}|p{0.2in}|p{0.2in}|p{0.18in}|p{0.21in}|p{0.2in}|p{0.2in}|}
    \hline 
	 & \multicolumn{4}{c|}{strided} & \multicolumn{4}{c|}{others} \\
    \hline
	config. & BL & BL + stride & DLA & DLA + T1 & BL & BL + stride & DLA & DLA + T1 \\
    \hline
	mean	   & 12.4 & 8.4 & 5.9 & 2.1 & 7.4 & 6.9 & 6.1 & 4.8 \\
    \hline 
	median	   & 10.0 & 4.8 & 4.0 & 1.1 & 3.9 & 3.5 & 2.8 & 3.2 \\
    \hline 
  \end{tabular}
\end{center}
	\caption{L1 MPKI divided between strided accesses and non-strided accesses,
	corresponding to four different configurations. For brevity, only means and
	median are shown.}\label{tab:mpki} 
\end{table}

With offloading stride prefetching, we move the comparatively simpler task
of prefetching for certain strided accesses to a hardware FSM. This alone
reduces the skeleton size (from 66\% to 45\% on average, more on that
later), allowing LT to run faster, and in turn making it more likely to
succeed in other prefetches. To understand the effect, we compare three
different options of prefetching these strided accesses: a modified stride
prefetcher~\cite{fu.micro92},\footnote{The baseline microarchitecture
contains a BOP at L2. The stride prefetcher was an additional prefetcher
to L1 and chosen from among 8 prefetchers~\cite{fu.micro92, nesbit.hpca04,
srinath.hpca07, somogyi.isca06, ishii.jilp11, shevgoor.micro15, kim.micro16,
michaud.hpca16} for best performance. Additionally, it was tuned to further
improve performance and is configured with 32 strides with a prefetch degree
of 4 for maximum performance using training inputs.} the baseline DLA, and
R3-DLA. We show the mean and median L1 MPKI (misses per kilo instructions) for
strided and the remaining accesses in Table~\ref{tab:mpki}.

\begin{figure}[htb]
	\centering   
	\subfigure[]{\includegraphics[width=0.48\columnwidth]{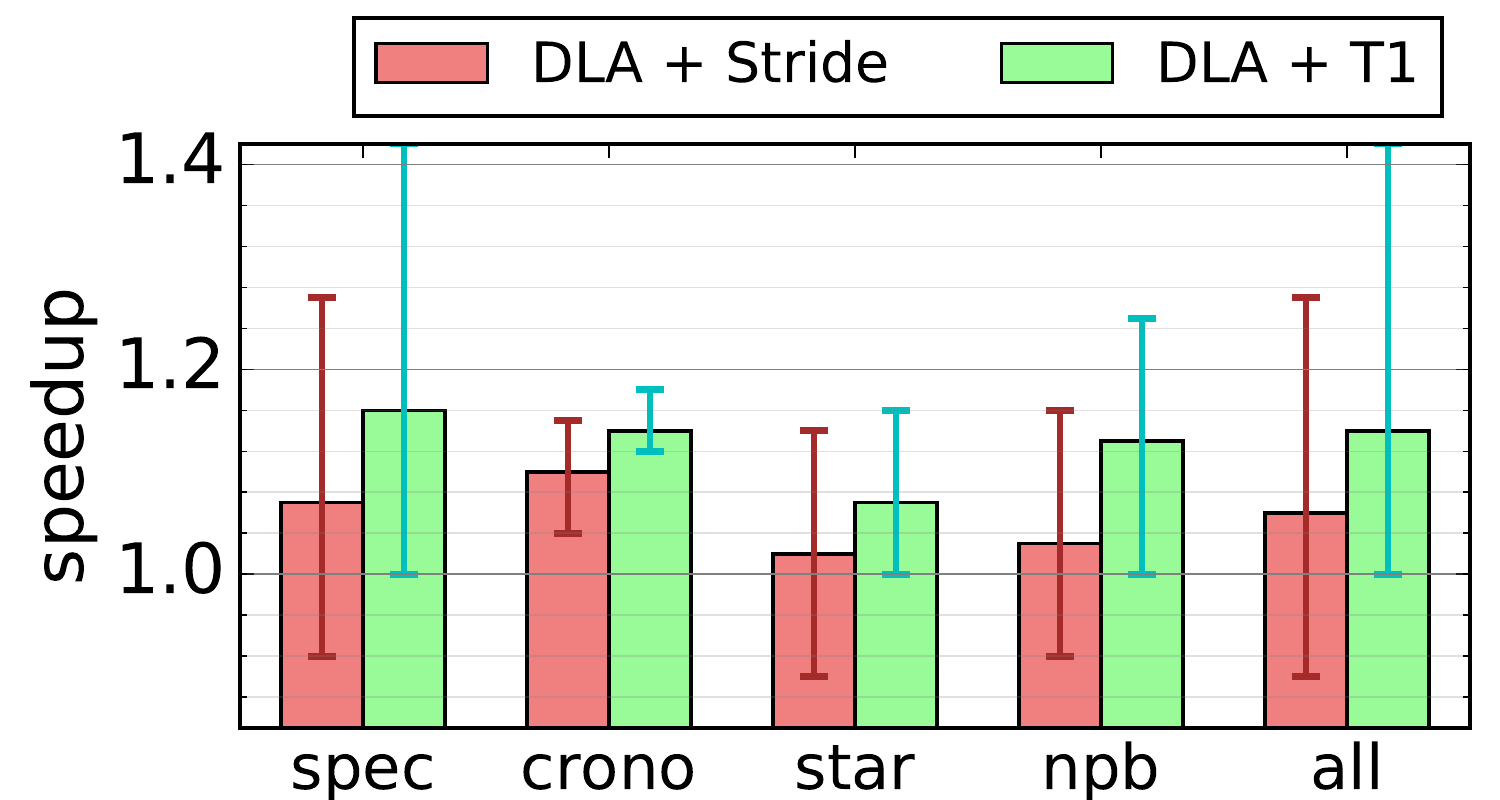}}
	\subfigure[]{\includegraphics[width=0.48\columnwidth]{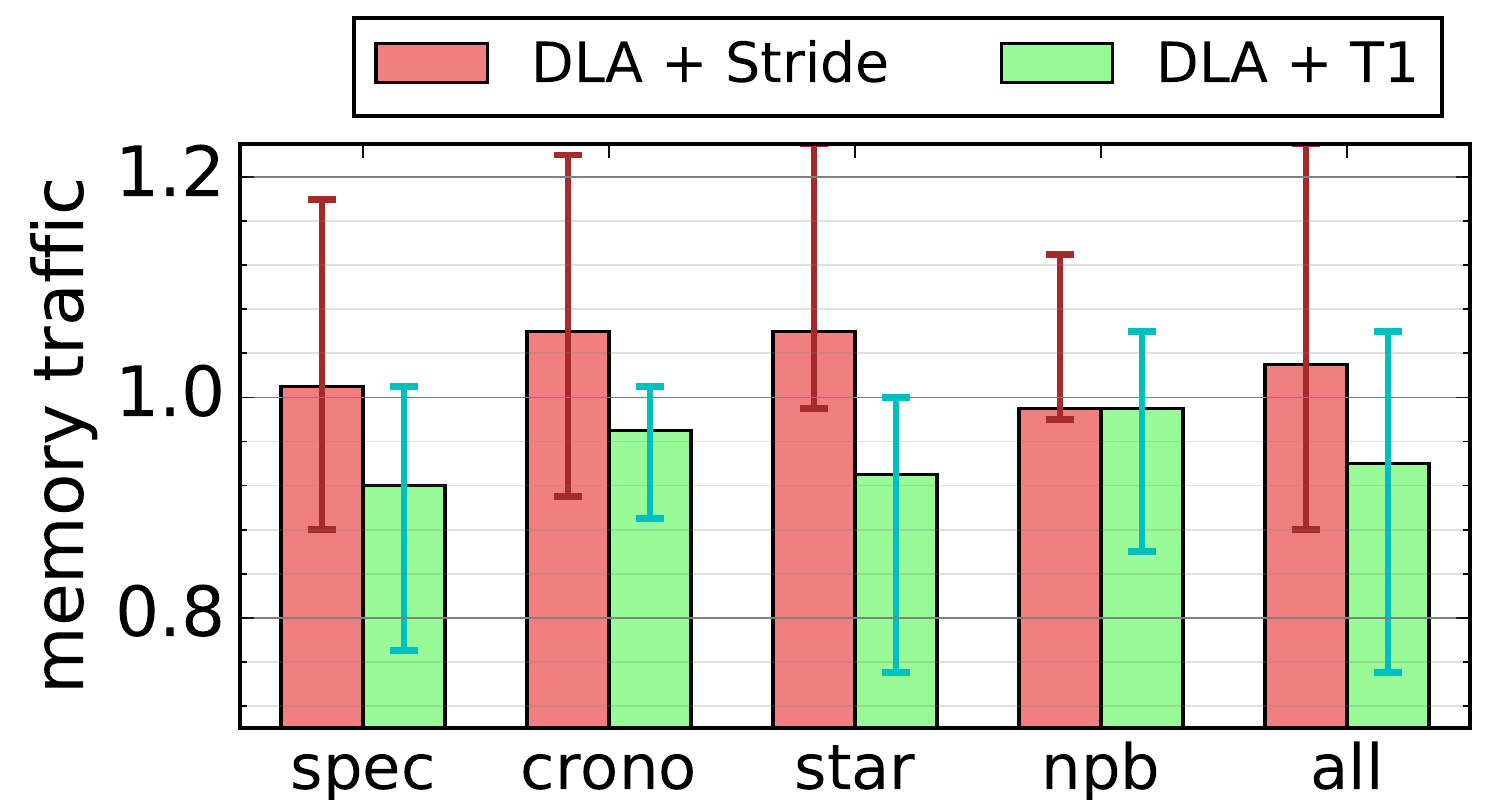}}
	\caption{Comparison of (a) speedup and (b) normalized memory traffic of
	two different configurations: DLA with a stride prefetcher and DLA with
	offloading (DLA+T1). \label{fig:ipc_t2}} 
\end{figure}

\begin{figure*}[b]
	\centering   
	\subfigure[]{\includegraphics[width=0.5\columnwidth]{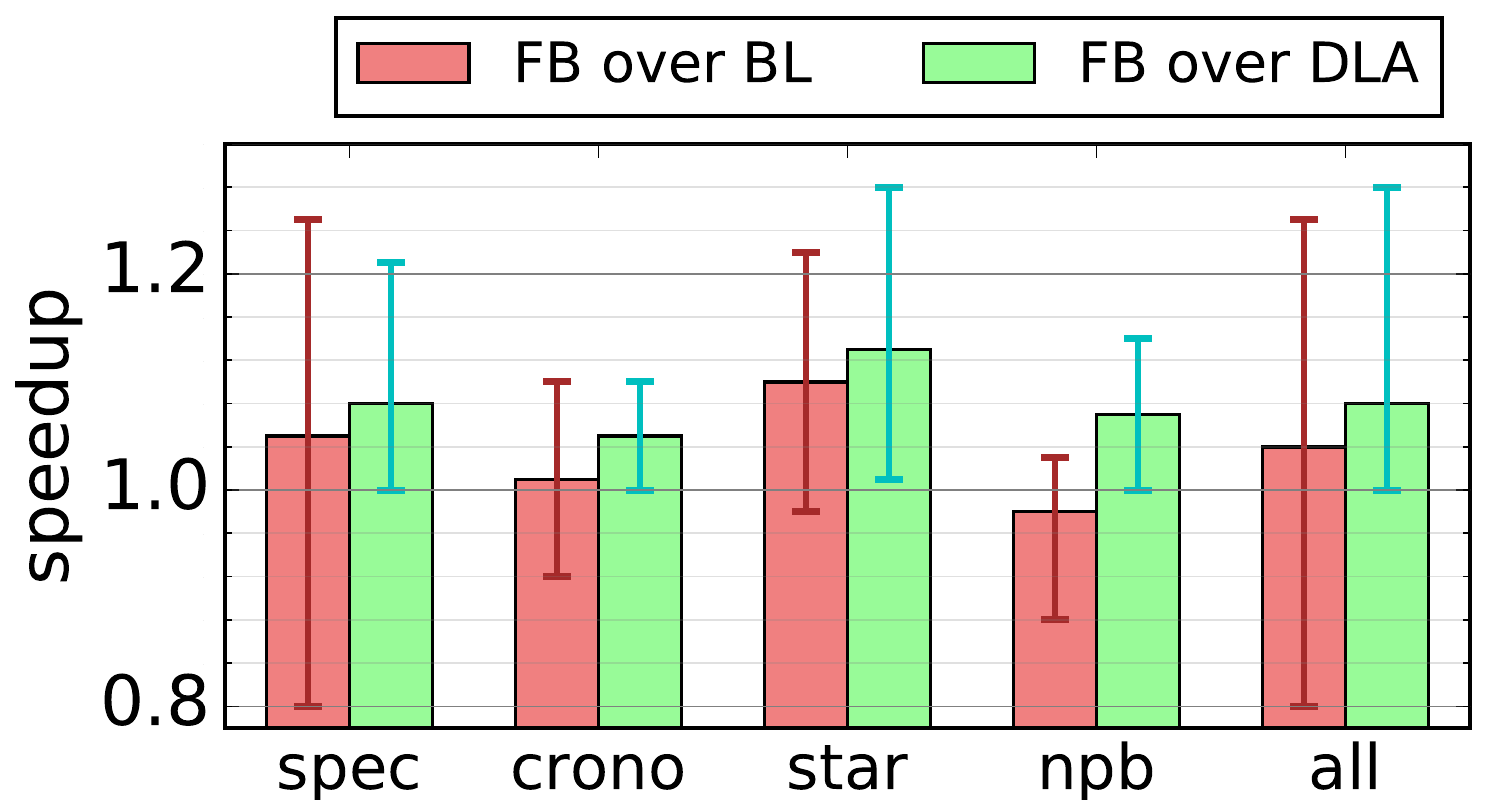}}
	\subfigure[]{\includegraphics[width=0.74\columnwidth]{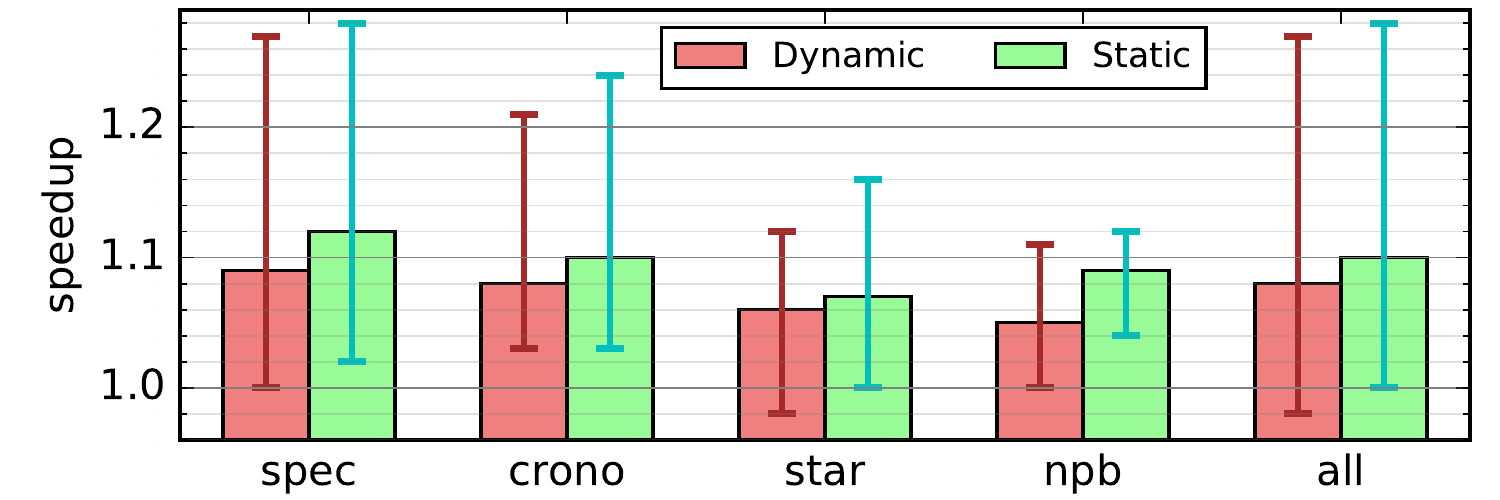}}
	\subfigure[]{\includegraphics[width=0.74\columnwidth]{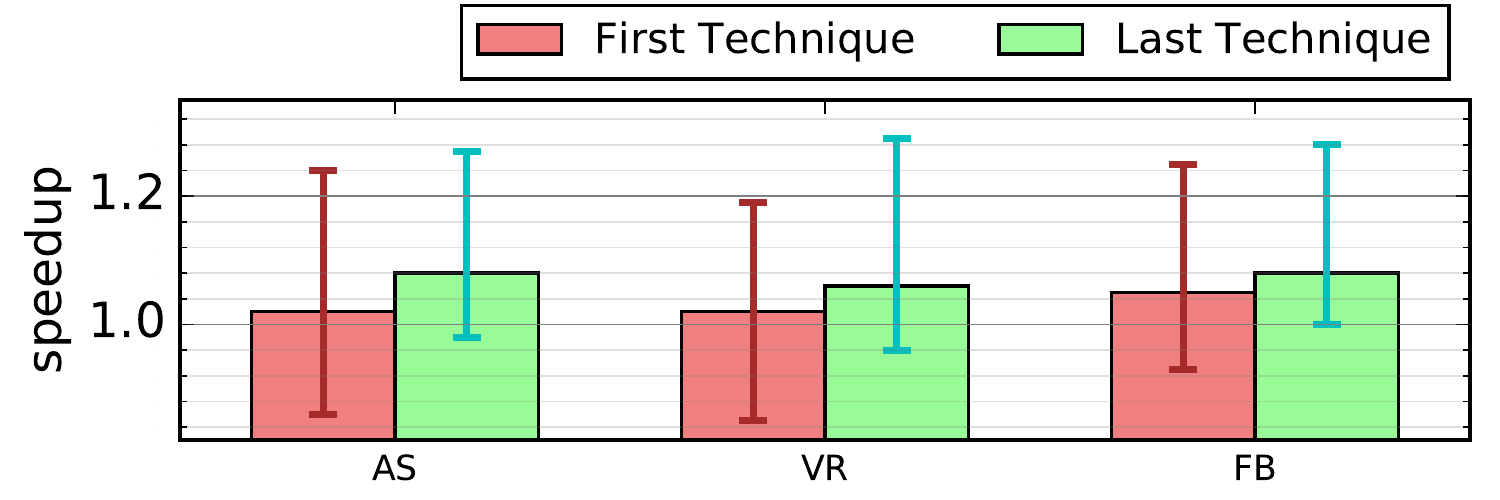}}
	\caption{(a) Comparison of performance gains obtained by a Fetch Buffer over BL
	system and over DLA system (b) The speedup differences between dynamic and
	static tuning (c) Speedup when an optimization is applied first or after other
	optimizations. \label{fig:fb_tuning_synergy}} 
\end{figure*}

We can imagine that T1 is far from perfect, requiring the loop to start a few
iterations before catching up. This is why there is still a non-trivial 2.1
MPKI remaining for strided accesses. But in comparison, both baseline DLA and
a hardware prefetcher are worse with 5.9 and 8.4 MPKI remaining respectively.
Additionally, the offloading improved DLA's ability to target non-strided
misses, reducing it from 6.1 to 4.8 MPKI on average. The medians show a
similar trend.

Figure~\ref{fig:ipc_t2} evaluates both performance and memory traffic metrics
among the various choices. For brevity, we only show the aggregate result as
the suite-wide average (represented by the bar) and range among individual
applications (represented by the I-beams).

First, we see that offloading works very well with DLA across all four
suites, achieving a geometric mean speedup of 1.14x over all
benchmarks. Second, this offloading arrangement is noticeably more
effective as well as more efficient than simply adding a hardware stride
prefetcher. In terms of performance, offloading never slows
down the system in any benchmarks and has a high mean speedup (compared
to 1.06x for adding a stride prefetcher). This is because the
T1 hardware does a much more limited and easier job than a conventional
stride prefetcher~\cite{chen.tc95}. This can be seen by the memory traffic
result shown in Figure~\ref{fig:ipc_t2}-b: the total memory traffic is lower
with adding T1 than with adding a stride prefetcher. Some of the extra
prefetches from the stride prefetcher are useless and create pollution, which
contributes to the lower performance.

\subsubsection{Reusing control flow information}

Using a fetch buffer to decouple fetch stage and decode stage is not a new
idea. The key point here is that DLA makes it far more effective due to the
much higher branch prediction accuracy. Figure~\ref{fig:fb_tuning_synergy}-a
shows the performance gains obtained by adding a fetch buffer over a baseline
system and a DLA system.

We see from the figure that the impact of a fetch buffer can be negative in
the baseline system. In fact, for NPB suite, the overall
effect is negative. When averaging over all applications, the benefit
is relatively small (4\% improvement). In contrast, when driven by the
highly-accurate prediction sequence from BOQ, the fetch buffer almost never
hurts and the benefit can be as high as 1.28x. Overall, the speedup due to
this addition is 1.08x.

\begin{figure}[h]
	\centering   
	\includegraphics[width=0.48\columnwidth]{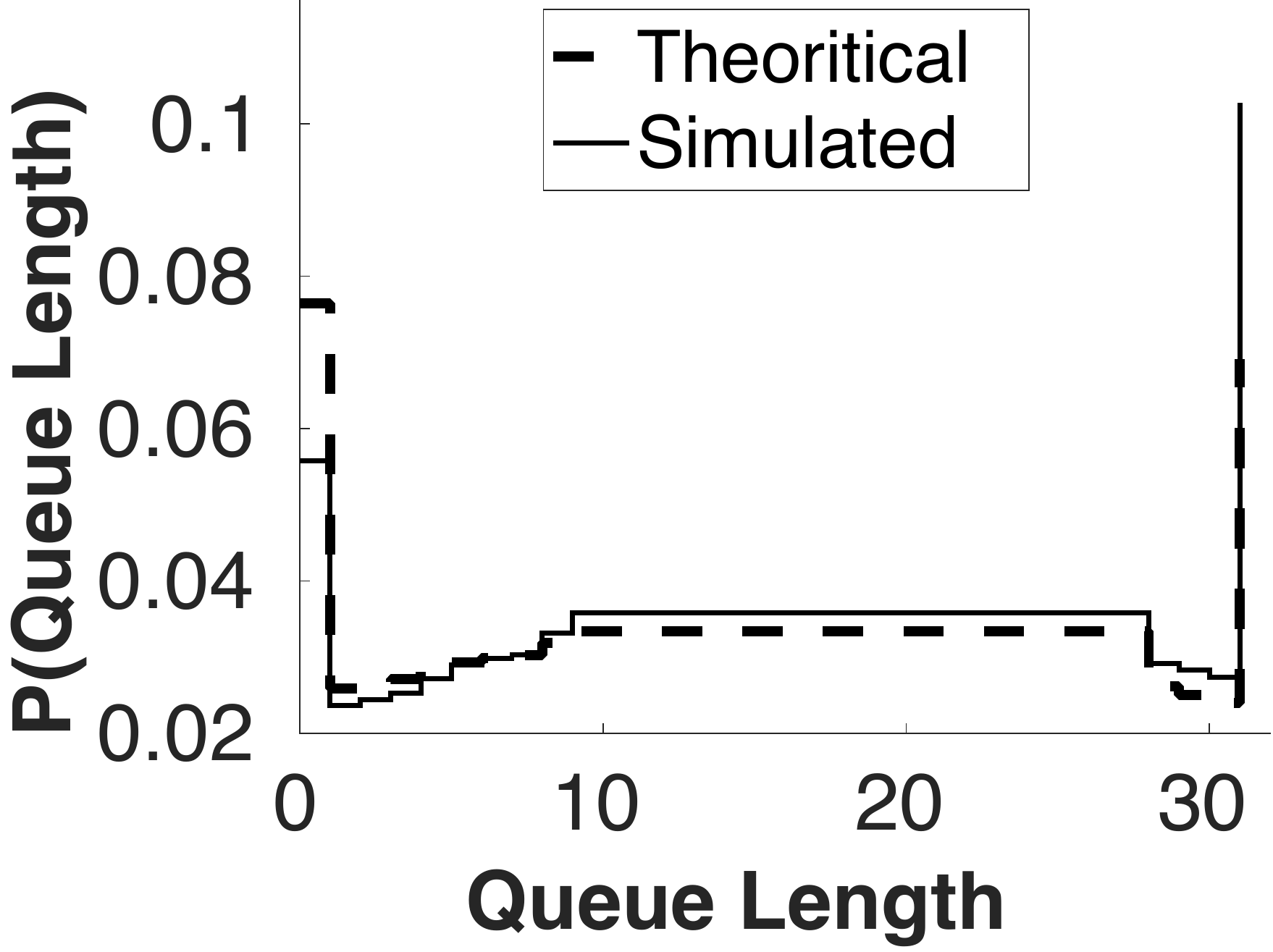}
	\caption{Comparison of theoretical and simulated probability distribution of
	fetch buffer queue length\label{fig:theory_fb}} 
\end{figure}

In our analysis of using the fetch buffer, we used a simplified probabilistic
approach. In Figure~\ref{fig:theory_fb}, we compare the theoretically derived
probability distribution of the queue length with that gathered from actual
simulation. We see that the general trend predicted by the theoretical analysis
agrees with simulation result rather well. 

\subsubsection{Re-cycling skeleton}
\label{ssec:atb_eval}

\begin{figure}[htb]
	\centering  
	\includegraphics[width=1\columnwidth]{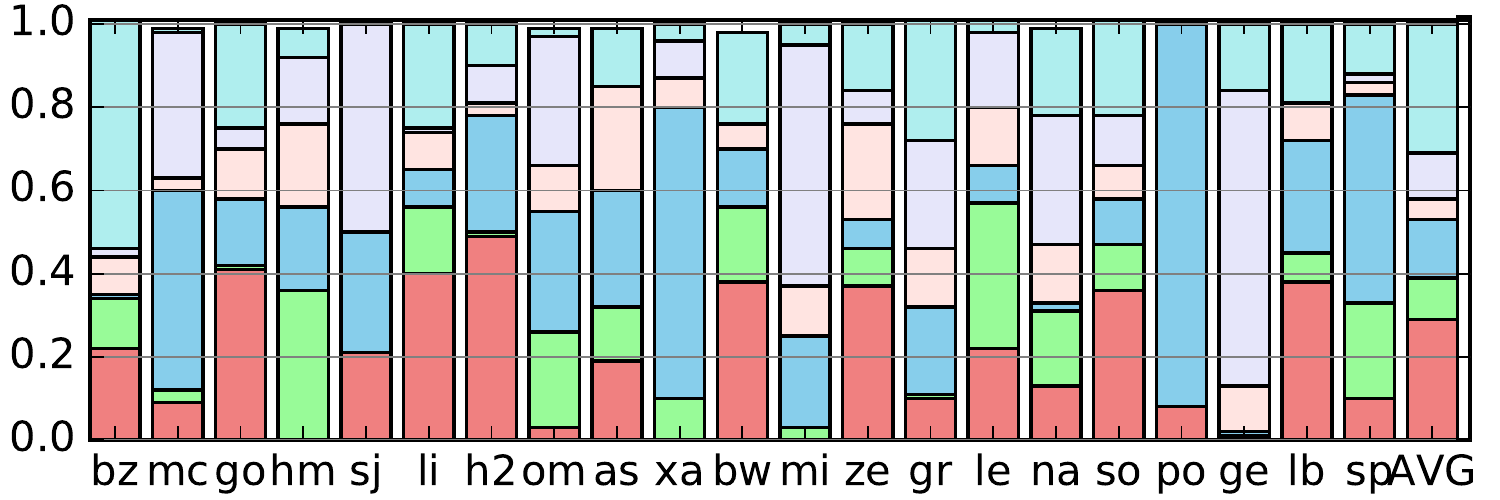}
	\caption{The distribution of skeleton versions chosen during online
	tuning.\label{fig:tuning}} 
\end{figure}

Re-cycling does not add new direct mechanisms for better look-ahead. It merely
searches the configuration space for a better solution. 
Figure~\ref{fig:tuning} shows the distribution of the skeleton being chosen
in re-cycling of skeleton. Each color shows a different configuration being
chosen. Although some simulation windows have a single choice for a major
portion of the window, all of them have chosen a number of different solutions
for different loops. This suggests that using simplistic heuristics to design
the default skeleton is unlikely to pick an optimal design for a particular
situation. Some dynamic tuning is perhaps necessary.

Our experiments show that re-cycling skeleton improves performance by about 1.08
on average and up to 1.27x as shown in Figure~\ref{fig:fb_tuning_synergy}-b.
The figure also compares the difference between dynamic on-line and static
off-line (using training inputs) tuning. We see that static tuning consistently
shows better result. This is partly due to the fact that in dynamic tuning, more
time is spent trying out suboptimal configurations. We note that in both
approaches, the tuning is done in a very crude way and in a coarse-grain manner.
This observation suggests that a more methodical, fine-grain tuning may be able
to further improve the performance of a DLA system.

\subsubsection{Synergy of individual optimizations}

While some optimizations proposed improve the speed of the look-ahead thread,
others extract more benefit from the look-ahead thread to improve the main
thread. There is an additional synergy when all these techniques are combined:
in a DLA system, the overall speed in a given phase is limited by the slower
of the two threads. So, if a technique speeds up only one thread, the system
performance will increase but only to the point where the other thread becomes
the new bottleneck. The rest of the benefit will only manifest when something
is done to improve the other thread. So, when multiple techniques are applied,
their combined benefit will usually be higher than implied by the benefit of
each individual technique measured in isolation. We show this visually in
Figure~\ref{fig:fb_tuning_synergy}-c.

In this experiment, we take the baseline DLA platform and measure the
performance impact of applying only one of the three techniques. We compare
that to applying the same technique last, that is, when the platform already
incorporated other techniques. We see that in all three cases, if we measure
the technique's benefit when it is applied as the first step, then none looks
especially promising averaging about 2-5\% gain. However, if we measure the
difference it makes as the last technique to be applied, the same technique
now appears to have a noticeably higher 6-8\% benefit. Thus, as we add more
optimizations to the design, more performance benefit may be unlocked.

%% file: env.tex
\subsection{Simulation Setup}
\label{ssec:setup}

For simulation purposes, we use Gem5~\cite{gem5} simulator to model
our proposed architecture. Our baseline is an aggressive out-of-order
pipeline with a Best Offset~\cite{michaud.hpca16} prefetcher (BOP) at
L2. This prefetcher is selected because in our experiments it provided
the best average performance gain among a group of 7 state-of-the-art
prefetchers~\cite{nesbit.hpca04, srinath.hpca07, somogyi.isca06,
ishii.jilp11, shevgoor.micro15, kim.micro16, michaud.hpca16} over all
application suites experimented. The
prefetcher is configured with 256 RR table entries and 52 offsets as described
in \cite{michaud.hpca16}. Additional technical details about the baseline are
provided in Table~\ref{table:config}. Unless otherwise mentioned, we use this
baseline configuration in all of our experiments. For DLA reboots, we add a
64 cycle delay to account for copying of the architectural registers from
the trailing thread to the look-ahead thread. Note that since the chances of
reboots are rare in DLA (0.6 on average in a 10k instruction window), their
impact on performance is minimal \eg increasing the reboot cost to 200 cycles
will degrade the overall performance of R3DLA by less than 2\%.

For comparison, we have also modeled a number of similar
approaches~\cite{kadjo.micro14, hashemi.micro16, purser.micro00} ranging
from earlier design of SlipStream to the recently proposed state-of-the-art
runahead execution scheme called Continuous Runahead Engine (CRE). Under our
baseline configuration (Table~\ref{table:config}), CRE outperforms other
designs. It generates its helper threads at runtime and executes them on a
custom processor located at the memory controller. We modified CRE's design
to prefetch data into L1 which on average provides higher overall performance
than just prefetching into LLC. Note that since we do not ignore any of the
applications in our evaluations and since our baseline configuration uses 3
levels of cache hierarchy with BOP as a L2 prefetcher, our performance numbers
for CRE appear different than the ones reported in \cite{hashemi.micro16}. For
similar configuration and applications, the average performance gain of CRE
on our platform is within 5\% of the ones reported in \cite{hashemi.micro16}.
In the case of \cite{garg.micro08}, the factors like memory model, improved
prefetcher/branch predictor and an overall aggressive baseline all contribute
to the variation in the reported performance benefits.

For CPU's energy consumption modeling, we use McPAT~\cite{mcpat} and assume a
22nm technology node. We modified McPAT to correctly model our proposed
architecture and additional hardware structures shown in Figure~\ref{fig:R3DLA}.
To compute main memory energy, we use DRAMPower~\cite{drampower}.

\begin{table}[htb]
\scriptsize
\centering
\begin{center}
  \begin{tabular}{|p{0.42in}|p{2.68in}|}
    \hline 
	Processing Node & 20-stage pipeline, out-of-order, 4-wide, 192 ROB, 
			          96 LSQ, 128INT/128FP PRF, 4INT/ 2MEM/ 4FP FUs, 
					  Tage SC-L Predictor (configured as the 256kBits
					  predictor described in \cite{seznec.jilp16}), 4K Entry
					  BTB, 32-entry RAS\\
	\hline
	Operating Points & 0.8V, 3.0GHz\\
   \hline
	L1 Caches & 32KB I-cache and 32KB D-cache, 4-way, 64B blocks, 3 ports, 1ns, 32 MSHRs, LRU \\
	\hline
	L2 Cache  & 256KB, 8-way, 64B blocks, 2 ports, 3ns, 32 MSHRs, LRU,
BOP~\cite{michaud.hpca16} \\
	\hline
	L3 Cache  & 2MB, 16-way, 64B blocks, 12ns, LRU \\
	\hline
	Main Memory & 4GB, DDR3 1600MHz, 1.5V, 2 channels, 2 ranks/channel, 8 banks/rank,
	t$_{RCD}$=13.75ns, t$_{RAS}$=35ns, t$_{FAW}$=30ns, t$_{WTR}$=7.5ns, t$_{RP}$=13.75ns \\
	\hline
	\hline
	\multicolumn{2}{c}{DLA Support} \\
	\hline
	BOQ& 512 entries (512x2 bits = 128B) \\
	\hline
	FQ& 128 entries (128x64 bits = 1KB) \\
	\hline
 	\hline
	\multicolumn{2}{c}{R3-DLA Support} \\
 	\hline
	T1 & 16 prefetching entries (512B) \\
 	\hline
	FB & 32 instructions (256B) \\
 	\hline
	VPT & 32 Entries (32x64 bits = 256B) (used by VR) \\
 	\hline
	LCT & 16 Entries (136B) (used by RC) \\
 	\hline
 	\hline
  \end{tabular}
\end{center}
	\caption{System configuration. \label{table:config}} 
\end{table}

We evaluate our proposal on a broad set of benchmark suites. In
addition to the SPEC2006~\cite{henning.can06} benchmark suite,
we use CRONO~\cite{masab.iiswc15} (a graph application suite),
STARBENCH~\cite{andersch.pars13} (embedded applications), and scientific
workloads from NAS Parallel Benchmarks (NPB). For SPEC2006, we use reference
inputs. For STARBENCH we use large inputs. NPB is simulated with C class of
workloads. For CRONO we use graph input data structures from google, amazon,
twitter, mathoverflow and california road-networks. All benchmarks are
compiled using gcc with -O3 option. To reduce simulation time, we use SimPoint
sampling methodology. To accurately capture all phases of the application we use
the SimPoint Tool~\cite{sherwood.asplos02} to generate five simpoints per
benchmark with 10 million instruction intervals. We warm up the caches for 100
million instructions before beginning each of the simpoint intervals. All the
simulation results are obtained from these simpoints.

%% file: conclusions.tex
\section{Conclusions}
\label{sec:conclusions}

Today's general-purpose applications continue to have significant levels
of implicit parallelism. However, data and instruction supply subsystem
presents significant barriers to exploiting this parallelism in a conventional
microarchitecture. Decoupled look-ahead systems are a potential solution.
In this paper, we have explored a number of optimizations to such an
architecture. They include \ding{172} reducing the look-ahead thread workload
by offloading simple prefetch pattern to a finite state machine; \ding{173}
reusing available values and control flow information to improve execution
and instruction supply to the main thread; and \ding{174} fine-tuning by
cycling through a number of pre-made skeletons. Each of these techniques
makes a seemingly limited contribution when applied in isolation. But
combined together, they improve the performance of a basic DLA system by
1.25x and achieves a speedup over a conventional architecture
with a state-of-the-art prefetcher by 1.4x on average. This
performance advantage differs from application to application and can be
as high as 2.24x, suggesting that if used selectively and
judiciously, an optimized R3-DLA system is already a high-performance solution
of exploiting the available implicit parallelism. Furthermore, analyses of the
system suggest the potential for further improvements.

%% file: appendix.tex
\section{Generating a basic DLA skeleton}
\label{sec:appskeleton}

The lead thread in DLA architecture executes a skeleton of the original
program binary. A skeleton includes all the control instructions and a subset
of memory instructions from the original program binary along with their
backward dependence chain. In this section we briefly describe the process of
generating this skeleton for DLA. Note that there is no correctness issue
with the skeleton as all information from LT is treated as predictions and is
thus fundamentally speculative.

\begin{figure}[h]
 	\centering  
	\includegraphics[width=3.2in]{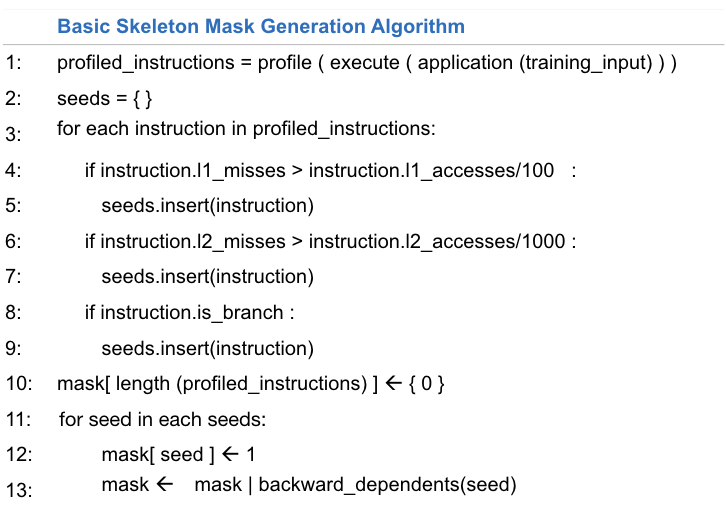}
	\caption{Pseudo code outlining the process of generating a basic skeleton
used by DLA.
	\label{fig:basic_skt_gen}} 
\end{figure}

The process first identifies the instructions for which the look ahead thread is
supposed to generate hints. These instructions, whom we call seeds, include all
the control instructions in the binary. If a runtime profiler is available, then
the program binary is executed with a training input to identify all the memory
instructions that have a higher probability of missing in the caches. In our
experiments, our runtime profiler selects all the memory instructions that have
more than 1\% chance of missing in L1 and/or more than 0.1\% chance of missing
in L2.  Once all the seeds are included, the skeleton generator identifies and
includes the backward dependents of each seed. While considering the backward
memory dependencies, the skeleton generator ignores all the store to load
dependencies if the store and the corresponding load are separated by more than
a 1000 static instructions. Note that not all of the memory dependencies can be
identified by the binary parser. However, the information generated by the
training run is sufficient to identify most of the memory dependencies. A pseudo
code for the skeleton generation process is presented in
Fig.~\ref{fig:basic_skt_gen}.

Note that our process of generating the skeleton is almost identical to the one
described in \cite{garg.micro08}, which includes more detailed discussions about
the choice of design parameters and additional optimizations.

%% file: app.tex
\section{Probabilistic Analysis of Fetch Buffer}
\label{sec:appfb}

We can measure the performance of a fetch unit by how many
\emph{fetch bubbles} it inserts into the pipeline down stream, \ie how many
more instructions the next stage (decode) can absorb but the fetch unit fails
to deliver. We make the simplifying assumption that the demand for instruction
from the next stage is an independent random variable from the status of
the queue. We use $D_j$ to denote the probability of the demand being $j$
instructions (thus $\sum_{j=0}^M D_j = 1$, $M$ being the decode width). We
use $Q_i$ to denote the probability of the queue containing $i$ instructions
(thus $\sum_{i=0}^N Q_i = 1$, $N$ being queue capacity). Then the expectation
of fetch bubbles can be calculated as follows. $$E(FB) = \sum_i^N \bigg(Q_i
\times \textstyle \sum_{j=i+1}^ND_j\times (j-i)\bigg)$$

In this calculation, the queue length
probability distribution is a function of capacity. We can empirically measure
it via simulations, but it is tedious and does not easily reveal insight.
Instead, we can analyze the queue as a simple Markov chain. The fetch queue
can be in one of $N+1$ states (holding between 0 and $N$ instructions).
We represent the probability distribution over these states as a vector.
$$Q\equiv [Q_0, Q_1, ..., Q_N]^T$$ 
To estimate the steady-state distribution ($Q^{ss}$) we need to calculate 
probabilities of state transitions. That, in turn, requires knowing the
probabilities of supplying and withdrawing instructions from the queue.
The calculation process is as follows.

\subsection{Change in queue length} Every cycle, the decode unit withdraws
instructions under a certain probability distribution ($D$ discussed above).
At the same time, the I-cache can supply new instructions following another
probability distribution ($S$). Convolving these two distributions will give
us a probability distribution of the \emph{change} in the queue length. This
probability distribution can be represented as a vector:
$$C\equiv [C_{-max_w}, ... , C_{max_d}]^T$$
where $C_\delta$ is the probability of a change of $\delta$ instructions and
$max_w$ and $max_d$ represent maximum number of instructions withdrawn or
deposited, respectively.

\subsection{Transition probability matrix} With probability distribution
$C$, we can now construct the transition matrix $\big[P_{i,j}\big]$ which
describes the conditional probability of having $i$ instructions in fetch
queue the next cycle when there are $j$ instructions in the current cycle.
Loosely speaking, the columns of matrix $P$ are merely vector $C$ shifted
appropriately such that $C_0$ aligns on the diagonal of the matrix. Since
the queue length can not be negative or higher than capacity, the boundary
elements ($P_{0,i}$ and $P_{N,i}$) absorb the portion of the vector left
outside the matrix. More formally, the elements are calculated as follows:
$$P_{i,j} = \begin{cases}
\sum_{k\leq i-j}C_k, & i=0; \\ 
\sum_{k\geq i-j}C_k, & i=N; \\ 
C_{(i-j)},  &otherwise.
\end{cases}$$

\subsection{Steady-state distribution} Given the probability transition matrix,
the probability distribution of the queue length in cycle $n+1$ can thus be
expressed as: $$Q^{(n+1)} = P\times Q^{(n)}$$ and thus steady-state solution
$Q^{ss} \equiv \lim_{n\to\infty}Q^{(n)}$ satisfies: $$Q^{ss} = P\times
Q^{ss}$$

That is to say $Q^{ss}$ is an eigenvector belonging to eigenvalue 1 of matrix
$P$. Since $P$ is a stochastic matrix, we know that there is a unique largest
eigenvalue of 1.\footnote{This is according to Perron-Frobenius theorem,
though strictly speaking, we require the matrix to be positive. We can imagine
substituting zero elements in $P$ with $\epsilon\to 0$. Another way of looking
at this problem is to diagonalize matrix $P$: $$\lim_{n\to\infty}Q^{(n)} =
\lim_{n\to\infty}(P^n) Q^{(0)}=X\lim_{n\to\infty}(\Lambda^n) X^{-1}Q^{(0)}
$$ Again, by Perron-Frobenius theorem, we know there is a unique largest
eigenvalue, so any initial probability distribution vector $Q^{(0)}$ will
lead to $Q^{ss}$.}

%opportunity to fetch more instructions and buffer them. We propose to have
%a queue between fetch stage and decode stage. The fetch stage of trailing
%thread continues to fetch as long as this queue is not full. Although, when
%the trailing thread is leading and using branch predictor (it does not use
%BOQ to predict its branches), this mechanism is turned off. This is not as
%beneficial in traditional architectures because of a relatively low accuracy
%of branch predictors.

\subsection{Case study} Here we use a 4-wide decode pipeline with a 16-wide
I-cache fetch width as a case study to analyze the impact of deeper fetch
decoupling. In our simplified model, the behavior of the program is reduced to
two probability distributions: $S$ and $D$. They are measured empirically.
In particular, we idealize the instruction fetch and measure the number of
instructions demanded to obtain $D$. Conversely, we idealize the backend of
the pipeline to measure the instruction supply probability distribution both
for an I-cache or a trace cache. Following the steps above, we obtain the
probability distribution of queue length ($Q$) with different capacities and
under both caches (shown in Fig.~\ref{fig:wabi}).

%\begin{figure}[h]
% 	\centering  
%	\subfigure[]{\includegraphics[width=2.8in]{FIGS/povray_pd.pdf}}
%	\subfigure[]{\includegraphics[width=2.8in]{FIGS/povray_efb.pdf}}
%	\caption{(a) The estimated probability distribution of queue length with
%	a queue capacity of 8 and 32 entries under both I-cache and
%	trace cache. (b) The expectation of fetch bubbles as the queue capacity
%	varies. In both cases, we show the result from one single application.
%	\label{fig:wabi}} 
%\end{figure}
%
%\begin{figure*}[htb]
% 	\centering  
%	%\includegraphics[width=2\columnwidth]{FIGS/ipc_1.pdf}
%	\includegraphics[width=2\columnwidth]{FIGS/r3dla_speedupspec_blBOL2_bar_stats.pdf}
%	\includegraphics[width=2\columnwidth]{FIGS/r3dla_speedupcr_blBOL2_bar_stats.pdf}
%	\includegraphics[width=2\columnwidth]{FIGS/r3dla_speedupnb_blBOL2_bar_stats.pdf}
%	\includegraphics[width=2\columnwidth]{FIGS/r3dla_speedupsb_blBOL2_bar_stats.pdf}
%	%\includegraphics[width=2\columnwidth]{FIGS/ipc_2.pdf}
%	\caption{Performance gain over an aggressive baseline with BO prefetcher at
%L2. For spec2k6, R3-DLA is 1.35x faster than baseline and 1.18x faster than DLA.
%For crono, it is 1.32x faster than baseline and 1.25x faster than DLA. For
%starbench, it is 1.33x faster than baseline and 1.29x faster than DLA. For npb,
%it is 1.35x faster than baseline and 1.19x faster than DLA.\label{fig:perf1}} 
%\end{figure*}